# Gibbs Sampling, Exponential Families and Orthogonal Polynomials[1]

**Persi Diaconis, Kshitij Khare and Laurent Saloff-Coste**

*Abstract.* We give families of examples where sharp rates of convergence to stationarity of the widely used Gibbs sampler are available. The examples involve standard exponential families and their conjugate priors. In each case, the transition operator is explicitly diagonalizable with classical orthogonal polynomials as eigenfunctions.

## 1. INTRODUCTION

The Gibbs sampler, also known as Glauber dynamics or the heat-bath algorithm, is a mainstay of scientific computing. It provides a way to draw samples from a multivariate probability density $f(x_1, x_2, \ldots, x_p)$, perhaps only known up to a normalizing constant, by a sequence of one-dimensional sampling problems. From $(X_1, \ldots, X_p)$ proceed to $(X'_1, X_2, \ldots, X_p)$, then $(X'_1, X'_2, X_3, \ldots, X_p), \ldots, (X'_1, X'_2, \ldots, X'_p)$ where at the $i$th stage, the coordinate is sampled from $f$ with the other coordinates fixed. This is one pass. Continuing gives a Markov chain $X, X', X'', \ldots$,

*Persi Diaconis is Professor, Department of Statistics, Sequoia Hall, 390 Serra Mall, Stanford University, Stanford, California 94305-4065, USA e-mail: diaconis@math.stanford.edu. Kshitij Khare is Graduate Student, Department of Statistics, Sequoia Hall, 390 Serra Mall, Stanford University, Stanford, California 94305-4065, USA e-mail: kdkhare@stanford.edu. Laurent Saloff-Coste is Professor, Department of Mathematics, Cornell University, Malott Hall, Ithaca, New York 14853-4201, USA e-mail: lsc@math.cornell.edu.*

which has $f$ as stationary density under mild conditions discussed in [4, 102].

The algorithm was introduced in 1963 by Glauber [49] to do simulations for Ising models, and independently by Turcin [103]. It is still a standard tool of statistical physics, both for practical simulation (e.g., [88]) and as a natural dynamics (e.g., [10]). The basic Dobrushin uniqueness theorem showing existence of Gibbs measures was proved based on this dynamics (e.g., [54]). It was introduced as a base for image analysis by Geman and Geman [46]. Statisticians began to employ the method for routine Bayesian computations following the works of Tanner and Wong [101] and Gelfand and Smith [45]. Textbook accounts, with many examples from biology and the social sciences, along with extensive references are in [47, 48, 78].

In any practical application of the Gibbs sampler, it is important to know how long to run the Markov chain until it has forgotten the original starting state. Indeed, some Markov chains are run on enormous state spaces (think of shuffling cards or image analysis). Do they take an enormous number of steps to reach stationarity? One way of dealing with these problems is to throw away an initial segment of the run. This leaves the question of how much to throw away. We call these questions running time analyses below.

Despite heroic efforts by the applied probability community, useful running time analyses for Gibbs sampler chains is still a major research effort. An overview of available tools and results is given at the end of this introduction. The main purpose of the present paper is to give families of two component examples where a sharp analysis is available. These may be used to compare and benchmark more robust techniques such as the Harris recurrence techniques in [64], or the spectral techniques in [2] and [107]. They may also serve as a base for the comparison techniques in [2, 26, 32].

Here is an example of our results. The following example was studied as a simple expository example in [16] and [78], page 132. Let

$$f_\theta(x) = \binom{n}{x} \theta^x (1-\theta)^{n-x},$$

$$\pi(d\theta) = \text{uniform on } (0,1), \quad x \in \{0,1,2,\ldots,n\}.$$

These define the bivariate Beta/Binomial density (uniform prior)

$$f(x,\theta) = \binom{n}{x} \theta^x (1-\theta)^{n-x}$$

with marginal density

$$m(x) = \int_0^1 f(x,\theta)\,d\theta$$
$$= \frac{1}{n+1}, \quad x \in \{0,1,2,\ldots,n\}.$$

The posterior density [with respect to the prior $\pi(d\theta)$] is given by

$$\pi(\theta|x) = \frac{f_\theta(x)}{m(x)}$$
$$= (n+1)\binom{n}{x}\theta^x(1-\theta)^{n-x}, \quad \theta \in (0,1).$$

The Gibbs sampler for $f(x,\theta)$ proceeds as follows:

- From $x$, draw $\theta'$ from $\text{Beta}(x+1, n-x+1)$.
- From $\theta'$, draw $x'$ from $\text{Binomial}(n,\theta')$.

The output is $(x',\theta')$. Let $\widetilde{K}(x,\theta;x',\theta')$ be the transition density for this chain. While $\widetilde{K}$ has $f(x,\theta)$ as stationary density, the $(\widetilde{K}, f)$ pair is not reversible (see below). This blocks straightforward use of spectral methods. Liu et al. [77] observed that the "$x$-chain" with kernel

$$k(x,x') = \int_0^1 f_\theta(x') \pi(\theta|x)\,d\theta$$
$$= \int_0^1 \frac{f_\theta(x) f_\theta(x')}{m(x)}\,d\theta$$

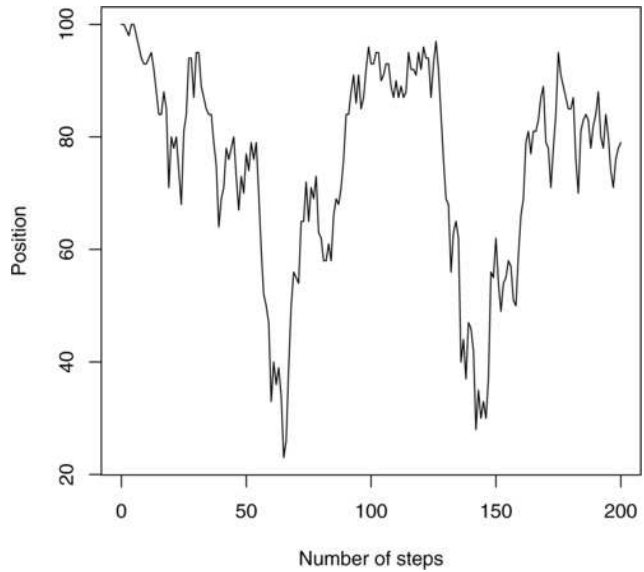

FIG. 1. *Simulation of the Beta/Binomial "x-chain" with $n = 100$.*

is reversible with stationary density $m(x)$ [i.e., $m(x) \cdot k(x,x') = m(x')k(x',x)$]. For the Beta/Binomial example

$$k(x,x') = \frac{n+1}{2n+1} \frac{\binom{n}{x}\binom{n}{x'}}{\binom{2n}{x+x'}},$$

(1.1)
$$0 \leq x, x' \leq n.$$

A simulation of the Beta/Binomial "$x$-chain" (1.1) with $n = 100$ is given in Figure 1. The initial position is 100 and we track the position of the Markov chain for the first 200 steps.

The proposition below gives an explicit diagonalization of the $x$-chain and sharp bounds for the bivariate chain [$\widetilde{K}_{n,\theta}^\ell$ denotes the density of the distribution of the bivariate chain after $\ell$ steps starting from $(n,\theta)$]. It shows that order $n$ steps are necessary and sufficient for convergence. That is, for sampling from the Markov chain $\widetilde{K}$ to simulate from the probability distribution $f$, a small (integer) multiple of $n$ steps suffices, while $\frac{n}{2}$ steps do not. The following bounds make this quantitative. The proof is given in Section 4. For example, when $n = 100$, after 200 steps the total variation distance (see below) to stationarity is less than 0.0192, while after 50 steps the total variation distance is greater than 0.1858, so the chain is far from equilibrium.

We simulate 3000 independent replicates of the Beta/Binomial "$x$-chain" (1.1) with $n = 100$, starting at the initial value 100. We provide histograms of the position of the "$x$-chain" after 50 steps and

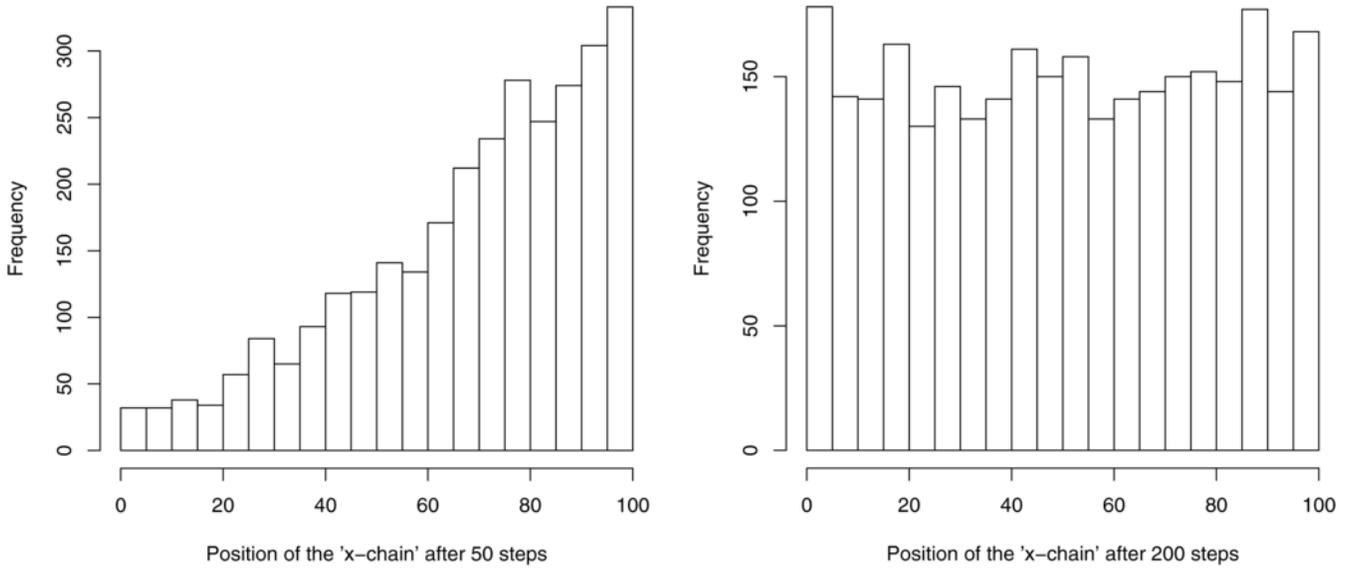

FIG. 2. *Histograms of the observations obtained from simulating* 3000 *independent replicates of the Beta/Binomial "x-chain" after* 50 *steps and* 200 *steps.*

200 steps in Figure 2. Under the stationary distribution (which is uniform on $\{0,1,\ldots,100\}$), one would expect roughly 150 observations in each block of the histogram. As expected, these histograms show that the empirical distribution of the position after 200 steps is close to the stationary distribution, while the empirical distribution of the position after 50 steps is quite different.

If $f, g$ are probability densities with respect to a $\sigma$-finite measure $\mu$, then the total variation distance between $f$ and $g$ is defined as

$$(1.2) \quad \|f - g\|_{\text{TV}} = \tfrac{1}{2} \int |f(x) - g(x)| \mu(dx).$$

PROPOSITION 1.1. *For the Beta/Binomial example with uniform prior, we have:*

(a) *The chain* (1.1) *has eigenvalues*

$$\beta_0 = 1,$$

$$\beta_j = \frac{n(n-1)\cdots(n-j+1)}{(n+2)(n+3)\cdots(n+j+1)}, \quad 1 \leq j \leq n.$$

*In particular,* $\beta_1 = 1 - 2/(n+2)$. *The eigenfunctions are the discrete Chebyshev polynomials [orthogonal polynomials for* $m(x) = 1/(n+1)$ *on* $\{0,\ldots,n\}$*].*

(b) *For the bivariate chain* $\widetilde{K}$, *for all* $\theta, n$ *and* $\ell$,

$$\frac{1}{2}\beta_1^\ell \leq \|\widetilde{K}_{n,\theta}^\ell - f\|_{\text{TV}} \leq \frac{\beta_1^{\ell-1/2}}{1 - \beta_1^{2\ell-1}}.$$

The calculations work because the operator with density $k(x, x')$ takes polynomials to polynomials. Our main results give classes of examples with the same explicit behavior. These include:

- $f_\theta(x)$ is one of the exponential families singled out by Morris [86, 87] (binomial, Poisson, negative binomial, normal, gamma, hyperbolic) with $\pi(\theta)$ the conjugate prior.
- $f_\theta(x) = g(x - \theta)$ is a location family with $\pi(\theta)$ conjugate and $g$ belongs to one of the six exponential families above.

Section 2 gives background. In Section 2.1 the Gibbs sampler is set up more carefully in both systematic and random scan versions. Relevant Markov chain tools are collected in Section 2.2. Exponential families and conjugate priors are reviewed in Section 2.3. The six families are described more carefully in Section 2.4 which calculates needed moments. A brief overview of orthogonal polynomials is in Section 2.5.

Section 3 is the heart of the paper. It breaks the operator with kernel $k(x, x')$ into two pieces: $T : L^2(m) \to L^2(\pi)$ defined by

$$Tg(\theta) = \int f_\theta(x) g(x) m(dx)$$

and its adjoint $T^*$. Then $k$ is the kernel of $T^*T$. Our analysis rests on a singular value decomposition of $T$. In our examples, $T$ takes orthogonal polynomials for $m(x)$ into orthogonal polynomials for $\pi(\theta)$. This leads to explicit computations and allows us to treat the random scan, $x$-chain and $\theta$-chain on an equal footing.

The $x$-chains and $\theta$-chains corresponding to three of the six classical exponential families are treated in Section 4. There are some surprises; while order $n$ steps are required for the Beta/Binomial example above, for the parallel Poisson/Gamma example, $\log n + c$ steps are necessary and sufficient. The six location chains are treated in Section 5 where some standard queuing models emerge (e.g., the $M/M/\infty$ queue). The final section points to other examples with polynomial eigenfunctions and other methods for studying present examples. Our examples are just illustrative. It is easy to sample from any of the families $f(x,\theta)$ directly. Further, we do not see how to carry our techniques over to higher component problems. We further point out that in routine Bayesian use of the Gibbs sampler, the distribution we wish to sample from is typically a posterior distribution of the parameters. Nevertheless, our two component problems are easy-to-understand standard statistical examples.

Basic convergence properties of the Gibbs sampler can be found in [4, 102]. Explicit rates of convergence appear in [95, 96]. These lean on Harris recurrence and require a drift condition of type $E(V(X_1)|X_0 = x) \leq aV(x) + b$ for all $x$. Also required are a minorization condition of the form $k(x,x') \geq \varepsilon q(x')$ for $\varepsilon > 0$, some probability density $q$, and all $x$ with $V(x) \leq d$. Here $d$ is fixed with $d \geq b/(1+a)$. Rosenthal [95] then gives explicit upper bounds and shows these are sometimes practically relevant for natural statistical examples. His paper [97] is a nice expository account. Finding useful $V$ and $q$ is currently a matter of art. For example, a group of graduate students tried to use these techniques in the Beta/Binomial example treated above and found it difficult to make choices giving useful results. This led to the present paper.

A marvelous expository account of this set of techniques with many examples and an extensive literature review is given by Jones and Hobert in [64]. In their main example an explicit eigenfunction was available for $V$; our Gamma/Gamma examples below generalize this. Further, "practically relevant" examples with useful analyses of Markov chains for stationary distributions which are difficult to sample directly from are in [65, 81, 98]. Some sharpenings of the Harris recurrence techniques are in [9] which also makes useful connections with classical renewal theory.

The Markov chains studied in this paper generate stochastic processes which have the marginal as stationary distribution. These chains have been used in a variety of applied modeling contexts in [80] and [89, 90, 91]. The present paper presents some new tools for these models. A related family of Markov chains and analyses have been introduced by Eaton [33, 34, 35] to study admissibility of formal Bayes estimators. The process is a two-component Gibbs sampler, as above, with $\pi$ an improper prior having an almost surely proper posterior. Under regularity assumptions, Eaton shows that recurrence of the $\theta$-chain implies the admissibility of certain formal Bayes estimators corresponding to $\pi$. Hobert and Robert [61] have developed this research, introducing the $x$-chain as a useful tool. Spectral techniques have proved to be a useful adjunct to Foster-type criteria for studying recurrence. We hope to develop our theory in these directions.

The analyses carried out in this paper hinge critically on the existence of all marginal and conditional moments. These moments need not exist. Indeed, consider the geometric density $f_\theta(x) = \theta^x(1-\theta), 0 \leq x < \infty$, and put a Beta$(\alpha, \beta)$ prior on $\theta$, with $\alpha \geq 1$ being an integer. The marginal distribution of $x$ admits only $\alpha - 1$ moments. For $\alpha > 1$, the available moments are put to good use in [25].

## 2. BACKGROUND

This section gives needed background. The two-component Gibbs sampler is defined more carefully in Section 2.1. Bounds on convergence using eigenvalues are given in Section 2.2. Exponential families and conjugate priors are reviewed in Section 2.3. The six families with variance a quadratic function of the mean are treated in Section 2.4. Finally, a brief review of orthogonal polynomials is in Section 2.5.

### 2.1 Two-Component Gibbs Samplers

Let $(\mathcal{X}, \mathcal{F})$ be a measurable space equipped with a $\sigma$-finite measure $\mu$. Let $(\Theta, \mathcal{G})$ be a measurable space equipped with a probability measure $\pi(d\theta)$. In many classical situations the prior is given by a density $g(\theta)$ with respect to the Lebesgue measure $d\theta$, and so $\pi(d\theta) = g(\theta)\, d\theta$. However, we also consider examples where the parameter $\theta$ is discrete, which cannot be described in the fashion mentioned above. Thus, we work with general prior probabilities $\pi(d\theta)$ throughout. Let $\{f_\theta(x)\}_{\theta \in \Theta}$ be a family of probability densities with respect to $\mu$. These define a probability measure on $\mathcal{X} \times \Theta$ via

$$P(A \times B) = \int_B \int_A f_\theta(x)\mu(dx)\pi(d\theta),$$
$$A \in \mathcal{F}, \ B \in \mathcal{G}.$$

The marginal density on $\mathcal{X}$ is

$$m(x) = \int_\Theta f_\theta(x)\pi(d\theta)$$

$$\left(\text{so } \int_{\mathcal{X}} m(x)\mu(dx) = 1\right).$$

We assume that $0 < m(x) < \infty$ for every $x \in \mathcal{X}$. The posterior density with respect to $\pi(d\theta)$ is given by

$$\pi(\theta|x) = f_\theta(x)/m(x).$$

The probability $P$ splits with respect to $m(dx) = m(x)\mu(dx)$ in the form

$$P(A \times B) = \int_A \int_B \pi(\theta|x)\pi(d\theta)m(dx),$$
$$A \in \mathcal{F}, \ B \in \mathcal{G}.$$

The systematic scan Gibbs sampler for drawing from the distribution $P$ proceeds as follows:

- Starting from $(x, \theta)$, first, draw $x'$ from $f_\theta(\cdot)$; second, draw $\theta'$ from $\pi(\cdot|x')$.
  The output is $(x', \theta')$. This generates a Markov chain $(x, \theta) \to (x', \theta') \to \cdots$ having kernel
  $$K(x, \theta; x', \theta') = f_\theta(x')f_{\theta'}(x')/m(x')$$
  with respect to $\mu(dx')\pi(d\theta')$. A slight variant exchanges the order of the draws:
- Starting from $(x, \theta)$, first, draw $\theta'$ from $\pi(\cdot|x)$; second, draw $x'$ from $f_{\theta'}(\cdot)$.
  The output is $(x', \theta')$. The corresponding Markov chain $(x, \theta) \to (x', \theta') \to \cdots$ has kernel
  $$\widetilde{K}(x, \theta; x', \theta') = f_{\theta'}(x)f_{\theta'}(x')/m(x)$$
  with respect to $\mu(dx')\pi(d\theta')$. Under mild conditions these two chains have stationary distribution $P$.

The "$x$-chain" [from $x$, draw $\theta'$ from $\pi(\theta'|x)$ and then $x'$ from $f_{\theta'}(x')$] has transition kernel

(2.1)
$$k(x, x') = \int_\Theta \pi(\theta|x)f_\theta(x')\pi(d\theta)$$
$$= \int_\Theta \frac{f_\theta(x)f_\theta(x')}{m(x)}\pi(d\theta).$$

Note that $\int k(x, x')\mu(dx') = 1$ so that $k(x, x')$ is a probability density with respect to $\mu$. Note further that $m(x)k(x, x') = m(x')k(x', x)$ so that the $x$-chain has $m(dx)$ as a stationary distribution.

The "$\theta$-chain" [from $\theta$, draw $x$ from $f_\theta(x)$ and then $\theta'$ from $\pi(\theta'|x)$] has transition density

(2.2)
$$k(\theta, \theta') = \int_{\mathcal{X}} f_\theta(x)\pi(\theta'|x)\mu(dx)$$
$$= \int_{\mathcal{X}} \frac{f_\theta(x)f_{\theta'}(x)}{m(x)}\mu(dx).$$

Note that $\int k(\theta, \theta')\pi(d\theta') = 1$ and that $k(\theta, \theta')$ has $\pi(d\theta)$ as reversing measure.

EXAMPLE (Poisson/Exponential). Let $\mathcal{X} = \{0, 1, 2, 3, \ldots\}$, $\mu(dx) =$ counting measure, $\Theta = (0, \infty)$, $f_\theta(x) = e^{-\theta}\theta^x/x!$. Take $\pi(d\theta) = e^{-\theta}\,d\theta$. Then $m(x) = \int_0^\infty \frac{e^{-\theta}\theta^x}{x!}e^{-\theta}\,d\theta = 1/2^{x+1}$. The posterior density with respect to $\pi(d\theta)$ is $\pi(\theta|x) = f_\theta(x)/m(x) = 2^{x+1}e^{-\theta}\theta^x/x!$. Finally, the $x$-chain has kernel

$$k(x, x') = \int_0^\infty \frac{2^{x+1}\theta^{x+x'}e^{-3\theta}}{x!x'!}\,d\theta$$
$$= \frac{2^{x+1}}{3^{x+x'+1}}\binom{x+x'}{x}, \quad 0 \leq x, x' < \infty,$$

whereas the $\theta$-chain has kernel

$$k(\theta, \theta') = 2e^{-\theta-\theta'}\sum_{x=0}^{\infty}\frac{(2\theta\theta')^x}{(x!)^2} = 2e^{-\theta-\theta'}I_0(\sqrt{8\theta\theta'})$$

with respect to $\pi(d\theta)$, where $I_0$ is the classical modified Bessel function; see Feller [44], Section 2.7, for background.

A second construction called the random scan chain is frequently used. From $(x, \theta)$, pick a coordinate at random and update it from the appropriate conditional distribution. More formally, for $g \in L^2(P)$

(2.3)
$$\bar{K}g(x, \theta) = \tfrac{1}{2}\int_\Theta g(x, \theta')\pi(\theta'|x)\pi(d\theta')$$
$$+ \tfrac{1}{2}\int_{\mathcal{X}} g(x', \theta)f_\theta(x')\mu(dx').$$

We note three things. First, $\bar{K}$ sends $L^2(P) \to L^2(P)$ and is reversible with respect to $P$. This is the usual reason for using random scans. Second, the right-hand side of (2.3) is the sum of a function of $x$ alone and a function of $\theta$ alone. That is, $\bar{K}: L^2(P) \to L^2(m) + L^2(\pi)$ [the range of $\bar{K}$ is contained in $L^2(m) + L^2(\pi)$]. Third, if $g \in (L^2(m) + L^2(\pi))^\perp$ [complement in $L^2(P)$], then $\bar{K}g = 0$ [Ker $\bar{K} \supseteq (L^2(m) + L^2(\pi))^\perp]$. Indeed, for any $h \in L^2(P)$, $\langle \bar{K}g, h \rangle_P = \int (\bar{K}g) h \, dP = \int (\bar{K}h) g \, dP = 0$. Thus $\bar{K}g = 0$. We diagonalize random scan chains in Section 3.

### 2.2 Bounds on Markov Chains

2.2.1 *General results.* We briefly recall well-known results that will be applied to either our two-component Gibbs sampler chains or the $x$- and $\theta$-chains. Suppose we are given a Markov chain described by its kernel $K(\xi, \xi')$ with respect to a measure $\tilde{\mu}(d\xi')$ [e.g., $\xi = (x, \theta)$, $\tilde{\mu}(d\xi) = \mu(dx)\pi(d\theta)$ for the two-component sampler, $\xi = \theta$, $\tilde{\mu}(d\theta) = \pi(d\theta)$ for the $\theta$-chain, etc.]. Suppose further that the chain has stationary measure $m(d\xi) = m(\xi)\tilde{\mu}(d\xi)$ and write

$$\hat{K}(\xi, \xi') = K(\xi, \xi')/m(\xi'),$$
$$\hat{K}^\ell_\xi(\xi') = \hat{K}^\ell(\xi, \xi') = K^\ell(\xi, \xi')/m(\xi')$$

for the kernel and iterated kernel of the chain with respect to the stationary measure $m(d\xi)$. We define the chi-square distance between the distribution of the chain started at $\xi$ after $\ell$ steps and its stationary measure by

$$\chi^2_\xi(\ell) = \int |\hat{K}^\ell_\xi(\xi') - 1|^2 m(d\xi')$$
$$= \int \frac{|K^\ell(\xi, \xi') - m(\xi')|^2}{m(\xi')} \tilde{\mu}(d\xi').$$

This quantity always yields an upper bound on the total variation distance

$$\|K^\ell_\xi - m\|_{\mathrm{TV}} = \tfrac{1}{2} \int |\hat{K}^\ell_\xi(\xi') - 1| m(d\xi')$$
$$= \tfrac{1}{2} \int |K^\ell(\xi, \xi') - m(\xi')| \tilde{\mu}(d\xi'),$$

namely,

(2.4) $\quad 4\|K^\ell_\xi - m\|^2_{\mathrm{TV}} \leq \chi^2_\xi(\ell).$

Our analysis will be based on eigenvalue decompositions. Let us first assume that we are given a function $\phi$ such that

$$K\phi(\xi) = \int K(\xi, \xi')\phi(\xi')\tilde{\mu}(d\xi') = \beta\phi(\xi),$$
$$m(\phi) = \int \phi(\xi) m(\xi) \tilde{\mu}(d\xi) = 0$$

for some (complex number) $\beta$. In words, $\phi$ is a generalized eigenfunction with eigenvalue $\beta$. We say "generalized" here because we have not assumed here that $\phi$ belongs to a specific $L^2$ space [we only assume we can compute $K\phi$ and $m(\phi)$]. The second condition [orthogonality to constants in $L^2(m)$] will be automatically satisfied when $|\beta| < 1$. Such an eigenfunction yields a simple lower bound on the convergence of the chain to its stationary measure.

LEMMA 2.1. *Referring to the notation above, assume that $\phi \in L^2(m(d\xi))$ and $\int |\phi|^2 \, dm = 1$. Then*

$$\chi^2_\xi(\ell) \geq |\phi(\xi)|^2 |\beta|^{2\ell}.$$

*Moreover, if $\phi$ is a bounded function, then*

$$\|K^\ell_\xi - m\|_{\mathrm{TV}} \geq \frac{|\phi(\xi)||\beta|^\ell}{2\|\phi\|_\infty}.$$

PROOF. This follows from the well-known results

(2.5) $\quad \chi^2_\xi(\ell) = \sup_{\|g\|_{2,m} \leq 1} \{|K^\ell_\xi(g) - m(g)|^2\}$

and

(2.6) $\quad \|K^\ell_\xi - m\|_{\mathrm{TV}} = \tfrac{1}{2} \sup_{\|g\|_\infty \leq 1} \{|K^\ell_\xi(g) - m(g)|\}.$

Here, $K^l_\xi(g)$ denotes the expectation of $g$ under the density $K^l(\xi, \cdot)$. For chi-square, use $g = \phi$ as a test function. For total variation use $g = \phi/\|\phi\|_\infty$ as a test function. More sophisticated lower bounds on total variation are based on the second moment method (e.g., [99, 106]). □

To obtain upper bounds on the chi-square distance, we need much stronger hypotheses. Namely, assume that $K$ is a self-adjoint operator on $L^2(m)$ and that $L^2(m)$ admits an orthonormal basis of real eigenfunctions $\varphi_i$ with real eigenvalues $\beta_i \geq 0$, $\beta_0 = 1$, $\varphi_0 \equiv 1$, $\beta_i \downarrow 0$ so that

$$\int \hat{K}(\xi, \xi')\varphi_i(\xi') m(d\xi') = \beta_i \varphi_i(\xi).$$

Assume further that $K$ acting on $L^2(m)$ is Hilbert–Schmidt (i.e., $\sum |\beta_i|^2 < \infty$). Then we have

$$\hat{K}^\ell(\xi, \xi') = \sum_i \beta_i^\ell \varphi_i(\xi) \varphi_i(\xi')$$

[convergence in $L^2(m \times m)$]

and

(2.7) $\quad \chi^2_\xi(\ell) = \sum_{i>0} \beta_i^{2\ell} \varphi_i^2(\xi).$

Useful references for this part of classical functional analysis are [1, 94].

2.2.2 *Application to the two-component Gibbs sampler.* All of the bounds in this paper are derived via the following route: bound $L^1$ by $L^2$ and use the explicit knowledge of eigenvalues and eigenfunctions to bound the sum in (2.7). This, however, does not apply directly to the two-component Gibbs sampler $K$ (or $\widetilde{K}$) because these chains are not reversible with respect to their stationary measure. Fortunately, the $x$-chain and the $\theta$-chain are reversible and their analysis yields bounds on the two-component chain thanks to the following elementary observation. The $x$-chain has kernel $k(x, x')$ with respect to the measure $\mu(dx)$. It will also be useful to have $\hat{k}(x, x') = k(x, x')/m(x')$, the kernel with respect to the probability $m(dx) = m(x)\mu(dx)$. For $\ell \geq 2$, we let $k_x^\ell(x') = k^\ell(x, x') = \int k(x,y) k^{\ell-1}(y, x')\mu(dy)$ denote the density [w.r.t. $\mu(dx)$] of the distribution of the $x$-chain after $l$ steps and set $\hat{k}_x^\ell(x') = \hat{k}^\ell(x, x') = \int \hat{k}(x,y)\hat{k}^{\ell-1}(y, x')m(dy)$ [the density w.r.t. $m(dx)$]. Also

$$\|g\|_{p,P} = \left(\int |g(\omega)|^p P(d\omega)\right)^{1/p} \quad \text{for } p \geq 1.$$

LEMMA 2.2. *Referring to the $K, \widetilde{K}$ two-component chains and $x$-chain introduced in Section 2.1, for any $p \in [1, \infty]$, we have*

$$\|(K_{x,\theta}^\ell/f) - 1\|_{p,P}^p \leq \int \|\hat{k}_z^{\ell-1} - 1\|_{p,m}^p f_\theta(z)\mu(dz)$$

$$\leq \sup_z \|\hat{k}_z^{\ell-1} - 1\|_{p,m}^p$$

*and*

$$\|(\widetilde{K}_{x,\theta}^\ell/f) - 1\|_{p,P}^p \leq \|\hat{k}_x^{\ell-1} - 1\|_{p,m}^p.$$

*Similarly, for the $\theta$-chain, we have*

$$\|(\widetilde{K}_{x,\theta}^\ell/f) - 1\|_{p,P}^p \leq \int \|k_\theta^{\ell-1} - 1\|_{p,\pi}^p \pi(\theta \mid x)\pi(d\theta)$$

$$\leq \sup_\theta \|k_\theta^{\ell-1} - 1\|_{p,\pi}^p$$

*and*

$$\|(K_{x,\theta}^l/f) - 1\|_{p,P}^p \leq \|k_\theta^{\ell-1} - 1\|_{p,\pi}^p.$$

PROOF. We only prove the results involving the $x$-chain. The rest is similar. Recall that the bivariate chain has transition density

$$K(x, \theta; x', \theta') = f_\theta(x') f_{\theta'}(x')/m(x').$$

By direct computation

$$K^\ell(x, \theta; x', \theta') = \int f_\theta(z) k^{\ell-1}(z, x') \frac{f_{\theta'}(x')}{m(x')} \mu(dz).$$

For the variant $\widetilde{K}$, the similar formula reads

$$\widetilde{K}^\ell(x, \theta; x', \theta') = \int k^{\ell-1}(x, z) \frac{f_{\theta'}(z)}{m(z)} f_{\theta'}(x')\mu(dz).$$

These two bivariate chains have stationary density $f(x, \theta) = f_\theta(x)$ with respect to the measure $\mu(dx) \cdot \pi(d\theta)$. So, we write

$$\frac{K^\ell(x, \theta; x', \theta')}{f(x', \theta')} - 1$$

$$= \int (\hat{k}^{\ell-1}(z, x') - 1) f_\theta(z)\mu(dz)$$

and

$$\frac{\widetilde{K}^\ell(x, \theta; x', \theta')}{f(x', \theta')} - 1$$

$$= \int (\hat{k}^{\ell-1}(x, z) - 1) f_{\theta'}(z)\mu(dz).$$

To prove the second inequality in the lemma (the proof of the first is similar), write

$$\|(\widetilde{K}_{x,\theta}^\ell/f) - 1\|_{p,P}^p$$

$$= \int \int \left|\int (\hat{k}^{\ell-1}(x, z) - 1) f_{\theta'}(z)\mu(dz)\right|^p$$

$$\cdot f_{\theta'}(x')\mu(dx')\pi(d\theta')$$

$$\leq \int \int \int |\hat{k}^{\ell-1}(x, z) - 1|^p$$

$$\cdot f_{\theta'}(z)\mu(dz) f_{\theta'}(x')\mu(dx')\pi(d\theta')$$

$$= \int |\hat{k}^{\ell-1}(x, z) - 1|^p m(z)\mu(dz)$$

$$= \int |\hat{k}^{\ell-1}(x, z) - 1|^p m(dz).$$

This gives the desired bound. □

To get lower bounds, we observe the following.

LEMMA 2.3. *Let $g$ be a function of $x$ only [abusing notation, $g(x, \theta) = g(x)$]. Then*

$$\widetilde{K}g(x, \theta) = \int k(x, x') g(x')\mu(dx').$$

*If instead, $g$ is a function of $\theta$ only, then*

$$Kg(x, \theta) = \int k(\theta, \theta') g(\theta')\pi(d\theta').$$

PROOF. Assume $g(x, \theta) = g(x)$. Then

$$\widetilde{K}g(x, \theta) = \int \int \frac{f_{\theta'}(x) f_{\theta'}(x')}{m(x)} g(x')\mu(dx')\pi(d\theta')$$

$$= \int k(x, x') g(x')\mu(dx').$$

The other case is similar. □

LEMMA 2.4. *Let $\chi^2_{x,\theta}(\ell)$ and $\widetilde{\chi}^2_{x,\theta}(\ell)$ be the chi-square distances after $\ell$ steps for the $K$-chain and the $\widetilde{K}$-chain, respectively, starting at $(x,\theta)$. Let $\chi^2_x(\ell)$, $\chi^2_\theta(\ell)$ be the chi-square distances for the $x$-chain (starting at $x$) and the $\theta$-chain (starting at $\theta$), respectively. Then we have*

$$\chi^2_\theta(\ell) \leq \chi^2_{x,\theta}(\ell) \leq \chi^2_\theta(\ell-1),$$

$$\|k^\ell_\theta - 1\|_{\mathrm{TV}} \leq \|K^\ell_{x,\theta} - f\|_{\mathrm{TV}} \leq \|k^{\ell-1}_\theta - 1\|_{\mathrm{TV}}$$

*and*

$$\chi^2_x(\ell) \leq \widetilde{\chi}^2_{x,\theta}(\ell) \leq \chi^2_x(\ell-1),$$

$$\|k^\ell_x - m\|_{\mathrm{TV}} \leq \|\widetilde{K}^\ell_{x,\theta} - f\|_{\mathrm{TV}} \leq \|k^{\ell-1}_x - m\|_{\mathrm{TV}}.$$

PROOF. This is immediate from (2.5)–(2.6) and Lemma 2.3. □

### 2.3 Exponential Families and Conjugate Priors

Three topics are covered in this section: exponential families, conjugate priors for exponential families and conjugate priors for location families.

2.3.1 *Exponential families.* Let $\mu$ be a $\sigma$-finite measure on the Borel sets of the real line $\mathbb{R}$. Define $\Theta = \{\theta \in \mathbb{R} : \int e^{x\theta} \mu(dx) < \infty\}$. Assume that $\Theta$ is nonempty and open. Hölder's inequality shows that $\Theta$ is an interval. For $\theta \in \Theta$, set

$$M(\theta) = \log \int e^{x\theta} \mu(dx),$$

$$f_\theta(x) = e^{x\theta - M(\theta)}.$$

The family of probability densities $\{f_\theta, \theta \in \Theta\}$ is the exponential family through $\mu$ in its "natural parametrization." Allowable differentiations yield the mean $m(\theta) = \int x f_\theta(x) \mu(dx) = M'(\theta)$ and the variance $\sigma^2(\theta) = M''(\theta)$.

Statisticians realized that many standard families can be put in such form so that properties can be studied in a unified way. Standard references for exponential families include [8, 11, 66, 73, 74].

EXAMPLE. Let $\mathcal{X} = \{0, 1, 2, 3, \ldots\}$, $\mu(x) = 1/x!$. Then $\Theta = \mathbb{R}$, and $M(\theta) = e^\theta$,

$$f_\theta(x) = \frac{e^{x\theta - e^\theta}}{x!}, \quad x = 0, 1, 2, \ldots.$$

This is the Poisson($\lambda$) distribution with $\lambda = e^\theta$.

This paper works with familiar exponential families. Many exotic families have been studied. See [12, 79]. These lead to interesting problems when studied in conjunction with the Gibbs sampler.

2.3.2 *Conjugate priors for exponential families.* With notation as above, fix $n_0 > 0$ and $x_0$ in the interior of the convex hull of the support of $\mu$. Define a prior density with respect to Lebesgue measure $d\theta$ by

$$\pi_{n_0, x_0}(d\theta) = z(n_0, x_0) e^{n_0 x_0 \theta - n_0 M(\theta)} d\theta,$$

where $z(n_0, x_0)$ is a normalizing constant shown to be positive and finite in Diaconis and Ylvisaker [29] which contains proofs of the assertions below. The posterior is

$$\pi(d\theta|x) = \pi_{n_0+1,(n_0 x_0 + x)/(n_0+1)}(d\theta).$$

Thus the family of conjugate priors is closed under sampling. This is sometimes used as the definition of conjugate prior. A central fact about conjugate priors is

(2.8) $$E(m(\theta)|x) = ax + b.$$

This linear expectation property characterizes conjugate priors for families where $\mu$ has infinite support. Section 3 shows that linear expectation implies that the associated chain defined at (2.1) always has an eigenfunction of the form $x - c$ with eigenvalue $a$, and $c$ equal to the mean of the marginal distribution.

Often, an exponential family is not parametrized by the natural parameter $\theta$, but in terms of the mean parameter $m(\theta)$. If we construct conjugate priors with respect to this parametrization, then (2.8) does not hold in general. However, for the six exponential families having quadratic variance function (discussed in Section 2.4 below), (2.8) holds even with the mean parametrization. In [19], this is shown to hold only for these six families. See [15, 55] for more on this.

EXAMPLE. For the Poisson example above the conjugate priors with respect to $\theta$ are of the form

$$z(n_0, x_0) e^{n_0 x_0 \theta - n_0 e^\theta} d\theta.$$

The mean parameter is $\lambda = e^\theta$. Since $d\theta = d\lambda/\lambda$, the priors transform to

$$z(n_0, x_0) \lambda^{n_0 x_0 - 1} e^{-n_0 \lambda} d\lambda.$$

The Poisson density parametrized by $\lambda$ is given by

$$f_\lambda(x) = \frac{e^{x \log \lambda - \lambda}}{x!}, \quad x = 0, 1, 2, \ldots.$$

Hence, the conjugate priors with respect to $\lambda$ are of the form

$$\tilde{z}(n_0, x_0)\lambda^{n_0 x_0} e^{-n_0 \lambda} d\lambda.$$

Here, the Jacobian of the transformation $\theta \to m(\theta)$ blends in with the rest of the prior so that both parametrizations lead to the usual Gamma priors for the Poisson density, which satisfy (2.8).

2.3.3 *Conjugate priors for location families.* Let $\mu$ be Lebesgue measure on $\mathbb{R}$ or counting measure on $\mathbb{N}$. In this section we consider random variables of the form $Y = \theta + \varepsilon$, with $\theta$ having density $\pi(\theta)$ and $\varepsilon$ having density $g(x)$ (both with respect to $\mu$). This can also be written as [densities w.r.t. $\mu(dx) \times \mu(d\theta)$]

$$f_\theta(x) = g(x - \theta),$$
$$f(x, \theta) = g(x - \theta)\pi(\theta).$$

In [30], a family of "conjugate priors" $\pi$ is suggested via posterior linearity. See [82] for further developments. The idea is to use the following well-known fact: If $X$ and $Y$ are independent random variables with finite means and the same distribution, then $E(X|X+Y) = (X+Y)/2$. More generally, if $X_r$ and $X_s$ are random variables which are independent with $X_r$ (resp. $X_s$) having the distribution of the sum of $r$ (resp. $s$) independent copies of the same random variable $Z$, then $E(X_r|X_r + X_s) = \frac{r}{r+s}(X_r + X_s)$. Here $r$ and $s$ may be taken as any positive real numbers if the underlying $Z$ is infinitely divisible.

With this notation, take $g$ as the density for $X_r$ and $\pi$ as the density for $X_s$ and call these a *conjugate location pair*. Then the marginal density $m(y)$ is the convolution of $g$ and $\pi$.

EXAMPLE. Let $g(x) = e^{-\lambda}\lambda^x/x!$ for $x \in \mathcal{X} = \{0, 1, 2, \ldots\}$. Take $\Theta = \mathcal{X}$ and let $\pi(\theta) = e^{-\eta}\eta^\theta/\theta!$. Then $m(x) = e^{-(\lambda+\eta)}(\lambda+\eta)^x/x!$ and

$$\pi(\theta|x) = \binom{x}{\theta}\left(\frac{\eta}{\lambda+\eta}\right)^\theta \left(\frac{\lambda}{\lambda+\eta}\right)^{x-\theta},$$
$$0 \leq \theta \leq x < \infty.$$

The Gibbs sampler (bivariate chain $K$) for this example becomes:

- From $x$, choose $\theta$ from Binomial$(x, \eta/(\eta+\lambda))$.
- From $\theta$, choose $X = \theta + \varepsilon$ with $\varepsilon \sim$ Poisson$(\lambda)$.

The $x$-chain may be represented as $X_{n+1} = S_{X_n} + \varepsilon_{n+1}$ with $S_k \sim$ Binomial$(k, \eta/(\eta+\lambda))$ and $\varepsilon \sim$ Poisson$(\lambda)$. This also represents the number of customers on service in an $M/M/\infty$ queue observed at discrete times: If this is $X_n$ at time $n$, then $S_{X_n}$ is the number served in the next time period and $\varepsilon_{n+1}$ is the number of unserved new arrivals. The explicit diagonalization of the $M/M/\infty$ chain, in continuous time, using Charlier polynomials appears in [3].

This same chain has yet a different interpretation: Let $f_\eta(j) = \binom{\eta}{j}p^j(1-p)^{\eta-j}$. Here $0 < p < 1$ is fixed and $\eta \in \{0, 1, 2, \ldots\}$ is a parameter. This model arises in underreporting problems where the true sample size is unknown. See [85]. Let $\eta$ have a Poisson$(\lambda)$ prior. The Gibbs sampler for the bivariate distribution $f(j, \eta) = \binom{\eta}{j}p^j(1-p)^{\eta-j}e^{-\lambda}\lambda^\eta/\eta!$ goes as follows:

- From $\eta$, choose $j$ from Bin$(\eta, p)$.
- From $j$, choose $\eta = j + \varepsilon$ with $\varepsilon \sim$ Poisson$(\lambda(1-p))$.

Up to a simple renaming of parameters, this is the same chain discussed above. Similar "translations" hold for any location problem where $\pi(\theta|x)$ has bounded range.

Note finally that there are natural statistical families and priors not of exponential form where the analysis works out neatly. The hypergeometric distribution for sampling from a finite population with a hypergeometric prior is developed in [28].

## 2.4 The Six Families

Morris [86, 87] has characterized exponential families where the variance $\sigma^2(\theta)$ is a quadratic function of the mean: $\sigma^2(\theta) = v_0 + v_1 m(\theta) + v_2 m^2(\theta)$. These six families have been characterized earlier by Meixner [83] in the development of a unified theory of orthogonal polynomials via generating functions. In [56] the same families are characterized in a regression context: For $X_i$ independent with a finite mean, $\bar{X} = \frac{1}{n}\sum X_i, S_n^2 = \frac{1}{n-1}\sum(X_i - \bar{X})^2$, one has

$$E(S_n^2|\bar{X} = \bar{x}) = a + b\bar{x} + c\bar{x}^2$$

if and only if the distribution of $X_i$ is one of the six families. In [39, 40], the six families are characterized by a link between orthogonal polynomials and martingales whereas [41, 92] makes a direct link to Lie theory. Finally, Consonni and Veronese [19] find the same six families in their study of conjugate priors: The conjugate priors in the natural parametrization given above transform into the same family in the mean parametrization only for the six families. Walter and Hamedani [105] construct orthogonal polynomials for the six families for use in empirical Bayes estimation. See also Pommeret [93].

Extensions are developed by Letac and Mora [76] and Casalis [15] who give excellent surveys of the literature. Still most useful, Morris [86, 87] gives a unified treatment of basic (and not so basic) properties such as moments, unbiased estimation, orthogonal polynomials and statistical properties. We give the six families in their usual parametrization along with the conjugate prior and formulae for the moments $E_\theta(X^k)$, $E_x(\theta^k)$ of $X$ and $\theta$ under $dP = f_\theta(x)\mu(dx)\pi(d\theta)$, given the value of the other. For each of these families, $E_\theta(X^k)$ and $E_x(\theta^k)$ are polynomials of degree $k$ in $\theta$ and $x$, respectively. We only specify the leading coefficients of these polynomials below. In fact, the leading coefficients are all we require for our analysis. In the conditional expectation formulae, $k$ is an integer running from 0 to $\infty$ unless specified otherwise. For $a \in \mathbb{R}$ and $n \in \mathbb{N} \cup \{0\}$, we define

$$(a)_n = a(a+1)\cdots(a+n-1)$$
$$\text{if } n \geq 1, (a)_0 = 1.$$

While some of the following calculations are standard and well known, others are not, and since the details enter our main theorems, we give complete statements.

*Binomial:* $\mathcal{X} = \{0, \ldots, n\}$, $\mu$ counting measure, $\Theta = (0,1)$:

$$f_\theta(x) = \binom{n}{x}\theta^x(1-\theta)^{n-x}, \quad 0 < \theta < 1,$$

$$\pi(d\theta) = \frac{\Gamma(\alpha+\beta)}{\Gamma(\alpha)\Gamma(\beta)}\theta^{\alpha-1}(1-\theta)^{\beta-1}\,d\theta,$$

$$0 < \alpha, \beta < \infty,$$

$$E_\theta(X^k) = (n-k+1)_k \theta^k + \sum_{j=0}^{k-1} a_j \theta^j,$$

$$0 \leq k \leq n,$$

$$E_x(\theta^k) = \frac{1}{(\alpha+\beta+n)_k}x^k + \sum_{j=0}^{k-1} b_j x^j.$$

*Poisson:* $\mathcal{X} = \{0, 1, 2, \ldots\}$, $\mu$ counting measure, $\Theta = (0, \infty)$:

$$f_\theta(x) = \frac{e^{-\theta}\theta^x}{x!}, \quad 0 < \theta < \infty,$$

$$\pi(d\theta) = \frac{\theta^{a-1}e^{-\theta/\alpha}}{\Gamma(a)\alpha^a}\,d\theta, \quad 0 < \alpha, a < \infty,$$

$$E_\theta(X^k) = \theta^k + \sum_{j=0}^{k-1} a_j \theta^j,$$

$$E_x(\theta^k) = \left(\frac{\alpha}{\alpha+1}\right)^k x^k + \sum_{j=0}^{k-1} b_j x^j.$$

*Negative Binomial:* $\mathcal{X} = \{0, 1, 2, \ldots\}$, $\mu$ counting measure, $\Theta = (0,1)$:

$$f_\theta(x) = \frac{\Gamma(x+r)}{\Gamma(r)x!}\theta^x(1-\theta)^r,$$

$$0 < \theta < 1, r > 0.$$

$$\pi(d\theta) = \frac{\Gamma(\alpha+\beta)}{\Gamma(\alpha)\Gamma(\beta)}\theta^{\alpha-1}(1-\theta)^{\beta-1}\,d\theta,$$

$$0 < \alpha, \beta < \infty.$$

$$E_\theta(X^k) = (r)_k\left(\frac{\theta}{1-\theta}\right)^k + \sum_{j=0}^{k-1} a_j\left(\frac{\theta}{1-\theta}\right)^j,$$

$$E_x\left(\left(\frac{\theta}{1-\theta}\right)^k\right) = (\beta+r-k)_k x^k + \sum_{j=0}^{k-1} b_j x^j,$$

$$k < \beta + r.$$

*Normal:* $\mathcal{X} = \Theta = \mathbb{R}$, $\mu$ Lebesgue measure:

$$f_\theta(x) = \frac{1}{\sqrt{2\pi\sigma^2}}e^{(-1/2)(x-\theta)^2/\sigma^2},$$

$$0 < \sigma^2 < \infty$$

$$\pi(d\theta) = \frac{1}{\sqrt{2\pi\tau^2}}e^{(-1/2)(\theta-v)^2/\tau^2}\,d\theta,$$

$$-\infty < v < \infty, 0 < \tau < \infty,$$

$$E_\theta(X^k) = \theta^k + \sum_{j=0}^{k-1} a_j \theta^j,$$

$$E_x(\theta^k) = \left(\frac{\tau^2}{\tau^2+\sigma^2}\right)^k x^k + \sum_{j=0}^{k-1} b_j x^j.$$

*Gamma:* $\mathcal{X} = \Theta = (0, \infty)$, $\mu$ Lebesgue measure:

$$f_\theta(x) = \frac{x^{a-1}e^{-x/\theta}}{\theta^a \Gamma(a)}, \quad 0 < a < \infty,$$

$$\pi(d\theta) = \frac{c^b \theta^{-(b+1)}e^{-c/\theta}}{\Gamma(b)}\,d\theta, \quad 0 < b, c < \infty,$$

$$E_\theta(X^k) = (a)_k \theta^k,$$

$$E_x(\theta^k) = (a+b-k)_k x^k + \sum_{j=0}^{k-1} b_j x^j,$$

$$0 \leq k < a+b.$$

*Hyperbolic:* $\mathcal{X} = \Theta = \mathbb{R}$, $\mu$ Lebesgue measure:

$$f_\theta(x) = \frac{2^{r-2}}{\pi r(1+\theta^2)^{r/2}}$$

$$\cdot e^{rx\tan^{-1}\theta}\beta\left(\frac{r}{2}+\frac{irx}{2},\frac{r}{2}-\frac{irx}{2}\right),$$
$$r > 0,$$
$$\text{where } \beta(a,b) = \frac{\Gamma(a)\Gamma(b)}{\Gamma(a+b)},$$
$$\pi(d\theta) = \frac{\Gamma(\rho/2-\rho\delta i/2)\Gamma(\rho/2+\rho\delta i/2)}{\Gamma(\rho/2)\Gamma(\rho/2-1/2)\sqrt{\pi}}$$
$$\cdot \frac{e^{\rho\delta\tan^{-1}\theta}}{(1+\theta^2)^{\rho/2}}\,d\theta,$$
$$-\infty < \delta < \infty,\ \rho \geq 1,$$
$$E_\theta(X^k) = k!\theta^k + \sum_{j=0}^{k-1} a_j \theta^j,$$
$$E_x(\theta^k) = \frac{1}{r^k(r+\rho-k-1)_k} x^k + \sum_{j=0}^{k-1} b_j x^j,$$
$$0 < k \leq r+\rho-1.$$

A unified way to prove the formulas involving $E_\theta(X^k)$ follows from Morris [87], (3.4). This says, for any of the six families with $m(\theta)$ the mean parameter and $p_k(x,m_0)$ the monic, orthogonal polynomials associated to the parameter $\theta_0$,

$$E_\theta(p_k(x,m_0)) = b_k(m(\theta)-m(\theta_0))^k,$$

where, if the family has variance function $\sigma^2(\theta) = v_2 m^2(\theta) + v_1 m(\theta) + v_0$,

$$b_k = \prod_{i=0}^{k-1}(1+iv_2).$$

For example, for the Binomial$(n,\theta)$ family, $m(\theta) = n\theta$, $\sigma^2(\theta) = n\theta(1-\theta)$, so $v_2 = -1/n$ and

$$E_\theta(p_k(x,m_0)) = \left\{\prod_{i=0}^{k-1}(n-i)\right\}(\theta-\theta_0)^k.$$

Comparing lead terms and using induction gives the first binomial entry. The rest are similar; the values of $v_2$ are $v_2(\text{Poisson}) = 0$, $v_2(\text{NB}) = 1/r$, $v_2(\text{Normal}) = 0$, $v_2(\text{Gamma}) = 1/r$, $v_2(\text{Hyperbolic}) = 1$. Presumably, there is a unified way to get the $E_x(\theta^k)$ entries, perhaps using [87], Theorem 5.4. This result shows that we get polynomials in $x$ but the lead coefficients do not come out as easily. At any rate, they all follow from elementary computations.

REMARKS. 1. The moment calculations above are transformed into a singular value decomposition and an explicit diagonalization of the univariate chains ($x$-chain, $\theta$-chain) in Section 3.

2. The first five families are very familiar, the sixth family less so. As one motivation, consider the generalized arc sine densities

$$f_\theta(y) = \frac{y^{\theta-1}(1-y)^{(1-\theta)-1}}{\Gamma(\theta)\Gamma(1-\theta)}, \quad 0 \leq y,\ \theta < 1.$$

Transform these to an exponential family via $x = \log(y/(1-y))$, $\eta = \pi\theta - \pi/2$. This has density

$$g_\eta(x) = \frac{e^{x\eta+\log(\cos\eta)}}{2\cosh((\pi/2)x)},$$
$$-\infty < x < \infty,\ -\frac{\pi}{2} < \eta < \frac{\pi}{2}.$$

The appearance of cosh explains the name hyperbolic. This density appears in [44], page 503, as an example of a density which is its own Fourier transform (like the normal). Many further references are in [37, 86, 87]. In particular, $g_0(x)$ is the density of $\frac{2}{\pi}\log|C|$ with $C$ standard Cauchy. The mean of $g_\eta(x)$ is $\tan(\eta) = \theta$. Parametrizing by the mean leads to the density shown with $r=1$. The average of $r$ independent copies of independent variates with this density gives the density with general $r$. The beta function is defined as usual; $\beta(a,b) = \Gamma(a)\Gamma(b)/\Gamma(a+b)$.

The conjugate prior for the mean parameter is of Pearson Type IV. When $\delta = 0$ this is a rescaled $t$ density. For general $\delta$ the family is called the skew $t$ in [37] which contains a wealth of information. Under the prior, the parameter $\theta$ has mean $\rho\delta/(\rho-2)$ and satisfies

$$(\rho-(k+2))E(\theta^{k+1}) = kE(\theta^{k-1}) + \rho\delta E(\theta^k),$$
$$1 \leq k < \rho-2.$$

This makes it simple to compute the $E_x(\theta^k)$ entry. Moments past $\rho$ are infinite.

The marginal distribution $m(x)$ can be computed in closed form. Using Stirling's formula in the form $|\Gamma(\sigma+it)| \sim \sqrt{2\pi}\ e^{-\pi|t|/2}|t|^{\sigma-1/2}$ as $|t| \uparrow \infty$, shows that $m(x)$ has tails asymptotic to $c/x^\rho$. It thus has only finitely many moments, so the $x$-chain must be studied by nonspectral methods. Of course, the additive version of our setup has moments of all order. The relevant orthogonal polynomials are Meixner–Pollaczek.

## 2.5 Some Background on Orthogonal Polynomials

A variety of orthogonal polynomials are used crucially in the following sections. While we usually just quote what we need from the extensive literature, this section describes a simple example. Perhaps the best introduction is in [18]. We will make frequent reference to [63] which is thorough and up-to-date. The classical account [100] contains much that is hard to find elsewhere. The on-line account [68] is very useful. For pointers to the literature on orthogonal polynomials and birth and death chains, see, for example, [104].

As an indication of what we need, consider the Beta/Binomial example with a general Beta$(\alpha, \beta)$ prior. Then the stationary distribution for the $x$-chain on $\mathcal{X} = \{0, 1, 2, \ldots, n\}$ is

$$m(x) = \binom{n}{x} \frac{(\alpha)_x (\beta)_{n-x}}{(\alpha + \beta)_n}$$

where

$$(a)_x = \frac{\Gamma(a+x)}{\Gamma(a)} = a(a+1) \cdots (a+x-1).$$

The choice $\alpha = \beta = 1$ yields the uniform distribution while $\alpha = \beta = 1/2$ yields the discrete arc-sine density from [43], Chapter 3,

$$m(x) = \frac{\binom{2x}{x} \binom{2n-2x}{n-x}}{2^{2n}}.$$

The orthogonal polynomials for $m$ are called Hahn polynomials [see (2.10) below]. They are developed in [63], Section 6.2, which refers to the very useful treatment of Karlin and McGregor [67]. The polynomials are given explicitly in [63], pages 178–179. Shifting parameters by 1 to make the classical notation match present notation, the orthogonal polynomials are

$$Q_j(x) = {}_3F_2\left(\begin{array}{c} -j, j+\alpha+\beta-1, -x \\ \alpha, -n \end{array} \bigg| 1 \right),$$

$$0 \le j \le n.$$

Here

(2.9)
$$_rF_s\left(\begin{array}{c} a_1 \cdots a_r \\ b_1 \cdots b_s \end{array} \bigg| z \right) = \sum_{\ell=0}^{\infty} \frac{(a_1 a_2 \cdots a_r)_\ell}{(b_1 b_2 \cdots b_s)_\ell} \frac{z^\ell}{\ell!}$$

$$\text{with } (a_1 \cdots a_r)_\ell = \prod_{i=1}^{r} (a_i)_\ell.$$

These polynomials satisfy

(2.10)
$$E_m(Q_j Q_\ell) = \frac{(\beta)_j (\alpha + \beta + j - 1)_{n+1}}{(\alpha + \beta + 2j - 1)}$$
$$\cdot \frac{j!(n-j)!}{(\alpha + \beta)_n (\alpha)_j n!} \delta_{jl}.$$

Thus they are orthogonal polynomials for $m$. When $\alpha = \beta = 1$, these become the discrete Chebyshev polynomials cited in Proposition 1.1. From our work in Section 2.2, we see we only need to know $Q_j(x_0)$ with $x_0$ the starting position. This is often available in closed form for special values, for example, for $x_0 = 0$ and $x_0 = n$,

(2.11)
$$Q_j(0) = 1,$$
$$Q_j(n) = \frac{(-\beta - j)_j}{(\alpha + 1)_j}, \quad 0 \le j \le n.$$

For general starting values, one may draw on the extensive work on uniform asymptotics; see, for example, [100], Chapter 8, or [5].

We note that [86], Section 8, gives an elegant self-contained development of orthogonal polynomials for the six families. Briefly, if $f_\theta(x) = e^{x\theta - M(\theta)}$ is the density, then

$$p_k(x, \theta) = \sigma^{2k} \left\{ \frac{d^k}{d^k m} f_\theta(x) \right\} \bigg/ f_\theta(x)$$

[derivatives with respect to the mean $m(\theta)$]. If $\sigma^2(\theta) = v_2 m^2(\theta) + v_1 m(\theta) + v_0$, then

$$E_\theta(p_n p_k) = \delta_{nk} a_k \sigma^{2k} \quad \text{with } a_k = k! \prod_{i=0}^{k-1}(1 + iv_2).$$

We also find need for orthogonal polynomials for the conjugate priors $\pi(\theta)$.

## 3. A SINGULAR VALUE DECOMPOSITION

The results of this section show that many of the Gibbs sampler Markov chains associated to the six families have polynomial eigenvectors, with explicitly known eigenvalues. This includes the $x$-chain, $\theta$-chain and the random scan chain. Analysis of these chains is in Sections 4 and 5. For a discussion of Markov operators related to orthogonal polynomials, see, for example, [6]. For closely related statistical literature, see [13] and the references therein.

Throughout, notation is as in Section 2.1. We have $\{f_\theta(x)\}_{\theta \in \Theta}$ a family of probability densities on the real line $\mathbb{R}$ with respect to a $\sigma$-finite measure $\mu(dx)$, for $\theta \in \Theta \subseteq \mathbb{R}$. Further, $\pi(d\theta)$ is a probability measure on $\Theta$. These define a joint probability $P$ on $\mathbb{R} \times \Theta$ with marginal density $m(x)$ [w.r.t. $\mu(dx)$] and posterior density [w.r.t. the prior $\pi(d\theta)$] given by $\pi(\theta|x) = f_\theta(x)/m(x)$. The densities do not have to come from exponential families in this section.

Let $c = \#\operatorname{supp} m(x)$. This may be finite or infinite. For simplicity, throughout this section, we assume $\operatorname{supp}(\pi)$ is infinite. Moreover, we make the following hypotheses:

(H1) For some $\alpha_1, \alpha_2 > 0$, $\int e^{\alpha_1|x|+\alpha_2|\theta|} P(dx, d\theta) < \infty$.

(H2) For $0 \leq k < c$, $E_\theta(X^k)$ is a polynomial in $\theta$ of degree $k$ with lead coefficient $\eta_k > 0$.

(H3) For $0 \leq k < \infty$, $E_x(\theta^k)$ is a polynomial in $x$ of degree $k$ with lead coefficient $\mu_k > 0$.

By (H1), $L^2(m(dx))$ admits a unique monic, orthogonal basis of polynomials $p_k$, $0 \leq k < c$, with $p_k$ of degree $k$. Also, $L^2(\pi(d\theta))$ admits a unique monic, orthogonal basis of polynomials $q_k$, $0 \leq k < \infty$, with $q_k$ of degree $k$. As usual, $\eta_0 = \mu_0 = 1$ and $p_0 \equiv q_0 \equiv 1$.

THEOREM 3.1. *Assume* (H1)–(H3). *Then:*

(a) *The $x$-chain* (2.1) *has eigenvalues* $\beta_k = \eta_k \mu_k$ *with eigenvectors* $p_k$, $0 \leq k < c$.

(b) *The $\theta$-chain* (2.2) *has eigenvalues* $\beta_k = \eta_k \mu_k$ *with eigenvectors* $q_k$ *for* $0 \leq k < c$, *and eigenvalues zero with eigenvectors* $q_k$ *for* $c \leq k < \infty$.

(c) *The random scan chain* (2.3) *has spectral decomposition given by*

*eigenvalues* $\frac{1}{2} \pm \frac{1}{2}\sqrt{\eta_k \mu_k}$,

*eigenvectors* $p_k(x) \pm \sqrt{\dfrac{\eta_k}{\mu_k}} q_k(\theta)$, $\quad 0 \leq k < c$,

*eigenvalues* $\frac{1}{2}$, *eigenvectors* $q_k$, $\quad c \leq k < \infty$.

The proof of Theorem 3.1 is in the Appendix.

REMARK. The theorem holds with obvious modification if $\#\operatorname{supp}(\pi) < \infty$. This occurs for binomial location problems. It will be used without further comment in Section 5. Further, the arguments work to give some eigenvalues with polynomial eigenvectors when only finitely many moments are finite.

## 4. EXPONENTIAL FAMILY EXAMPLES

This section carries out the analysis of the $x$- and $\theta$- chains for the Beta/Binomial, Poisson/Gamma and normal families. The $x$- and $\theta$-chains for the normal family are essentially the same. Hence, this section consists of five examples in all. For each, we set up the results for general parameter values and carry out the bounds in some natural special cases. The other three families are not amenable to this analysis due to lack of existence of all moments [which violates hypothesis (H1) in Section 3]. However, they are analyzed by probabilistic techniques such as coupling in [25].

### 4.1 Beta/Binomial

4.1.1 *The $x$-chain for the Beta/Binomial.* Fix $\alpha$, $\beta > 0$. On the state space $\mathcal{X} = \{0, 1, 2, \ldots, n\}$, let

$$
\begin{aligned}
k(x, x') &= \int_0^1 \binom{n}{x'} \theta^{\alpha+x+x'-1}(1-\theta)^{\beta+2n-(x+x')-1} \\
&\quad \cdot \frac{\Gamma(\alpha+\beta+n)\,d\theta}{\Gamma(\alpha+x)\Gamma(\beta+n-x)} \\
&= \binom{n}{x'} \frac{\Gamma(\alpha+\beta+n)\Gamma(\alpha+x+x')}{\Gamma(\alpha+x)\Gamma(\beta+n-x)} \\
&\quad \cdot \frac{\Gamma(\beta+2n-(x+x'))}{\Gamma(\alpha+\beta+2n)}.
\end{aligned}
\tag{4.1}
$$

When $\alpha = \beta = 1$ (uniform prior), $k(x, x')$ is given by (1.1). For general $\alpha, \beta$, the stationary distribution is the Beta/Binomial:

$$ m(x) = \binom{n}{x} \frac{(\alpha)_x (\beta)_{n-x}}{(\alpha+\beta)_n}, $$

where

$$ (a)_j = \frac{\Gamma(a+j)}{\Gamma(a)} = a(a+1)\cdots(a+j-1). $$

From our work in previous sections we obtain the following result.

PROPOSITION 4.1. *For $n = 1, 2, \ldots$, and $\alpha, \beta > 0$, the Beta/Binomial $x$-chain* (4.1) *has:*

(a) *Eigenvalues $\beta_0 = 1$ and $\beta_j = \dfrac{n(n-1)\cdots(n-j+1)}{(\alpha+\beta+n)_j}$, $1 \leq j \leq n$.*
(b) *Eigenfunctions $Q_j$, $0 \leq j \leq n$, the Hahn polynomials of Section* 2.5.
(c) *For any $\ell \geq 1$ and any starting state $x$,*

$$ \chi_x^2(\ell) = \sum_{i=1}^n \beta_i^{2\ell} Q_i^2(x) z_i $$

*where* $z_i = \dfrac{(\alpha+\beta+2i-1)(\alpha+\beta)_n (\alpha)_i}{(\beta)_i (\alpha+\beta+i-1)_{n+1}} \binom{n}{i}.$

We now specialize this to $\alpha = \beta = 1$ and prove the bounds announced in Proposition 1.1.

PROOF OF PROPOSITION 1.1. From (a), $\beta_i = \frac{n(n-1)\cdots(n-i+1)}{(n+2)(n+3)\cdots(n+i+1)}$. From (2.11), $Q_i^2(n) = 1$. By elementary manipulations, $z_i = \beta_i(2i+1)$. Thus

$$\chi_n^2(\ell) = \sum_{y=0}^{n} \frac{(k^\ell(n,y) - m(y))^2}{m(y)}$$
$$= \sum_{i=1}^{n} \beta_i^{2\ell+1}(2i+1).$$

We may bound $\beta_i \le \beta_1^i = (1 - \frac{2}{n+2})^i$, and so

$$\chi_n^2(\ell) = \sum_{i=1}^{n} \beta_i^{2\ell+1}(2i+1) \le \sum_{i=1}^{n} \beta_1^{i(2\ell+1)}(2i+1).$$

Using $\sum_1^\infty x^i = x/(1-x)$, $\sum_1^\infty ix^i = x/(1-x)^2$, we obtain

$$3\beta_1^{2\ell+1} \le \chi_n^2(\ell) \le \frac{3\beta_1^{2\ell+1}}{(1-\beta_1^{2\ell+1})^2}.$$

By Lemma 2.4, this gives (for the $\widetilde{K}$ chain)

$$3\beta_1^{2\ell+1} \le \widetilde{\chi}_{n,\theta}^2(\ell) \le \frac{3\beta_1^{2\ell-1}}{(1-\beta_1^{2\ell-1})^2}.$$

The upper bound in total variation follows from (2.4). For a lower bound in total variation, use the eigenfunction $\varphi_1(x) = x - \frac{n}{2}$. This is maximized at $x = n$ and the lower bound follows from Lemma 2.1. □

REMARK. Essentially, the same results hold for any Beta($\alpha, \beta$) prior in the sense that, for fixed $\alpha, \beta$, starting at $n$, order $n$ steps are necessary and sufficient for convergence. The computation gets more involved if one starts from a different point than $n$. Mizan Rahman and Mourad Ismail have shown us how to evaluate the Hahn polynomials at $n/2$ when $n$ is even and $\alpha = \beta$ using [63], (1.4.12) (the odd-degree Hahn polynomials vanish at $n/2$ and the three-term recurrence then easily yields the values of the even-degree Hahn polynomials at $n/2$). See Section 5.1 for a closely related example.

4.1.2 *The $\theta$-chain for the Beta/Binomial.* Fix $\alpha, \beta > 0$. On the state space $[0, 1]$, let

$$k(\theta, \theta') = \sum_{j=0}^{n} \binom{n}{j} \theta^j (1-\theta)^{n-j}$$

(4.2)
$$\cdot \frac{\Gamma(\alpha + \beta + n)}{\Gamma(\alpha + j)\Gamma(\beta + n - j)}$$
$$\cdot (\theta')^{\alpha + j - 1}(1 - \theta')^{\beta + n - j - 1}.$$

This is a transition density with respect to Lebesgue measure $d\theta'$ on $[0, 1]$. It has stationary density

$$\pi(d\theta) = \frac{\Gamma(\alpha + \beta)}{\Gamma(\alpha)\Gamma(\beta)} \theta^{\alpha - 1}(1 - \theta)^{\beta - 1} d\theta.$$

REMARK. In this specific example, the prior $\pi(d\theta)$ has a density $g(\theta) = \frac{\Gamma(\alpha+\beta)}{\Gamma(\alpha)\Gamma(\beta)}\theta^{\alpha-1}(1-\theta)^{\beta-1}$ with respect to the Lebesgue measure $d\theta$. For ease of exposition, we deviate from the general treatment in Section 2.1, where $k(\theta, \theta')$ is a transition density with respect to $\pi(d\theta')$. Instead, we absorb $g(\theta')$ in the transition density, so that $k(\theta, \theta')$ is a transition density with respect to the Lebesgue measure $d\theta'$.

The relevant orthogonal polynomials are Jacobi polynomials $P_i^{a,b}$, $a = \alpha - 1$, $b = \beta - 1$, given on $[-1, 1]$ in standard literature [68], 1.8. We make the change of variables $\theta = \frac{1-x}{2}$ and write $p_i(\theta) = P_i^{\alpha-1,\beta-1}(1-2\theta)$. Then, we have

(4.3) $$\int_0^1 p_j(\theta)p_k(\theta)\pi(\theta)\, d\theta = z_j^{-1}\delta_{jk},$$

where

$$z_j = \frac{(2j+\alpha+\beta-1)\Gamma(\alpha)\Gamma(\beta)\Gamma(j+\alpha+\beta-1)j!}{\Gamma(\alpha+\beta)\Gamma(j+\alpha)\Gamma(j+\beta)}.$$

PROPOSITION 4.2. *For $\alpha, \beta > 0$, the $\theta$-chain for the Beta/Binomial* (4.2) *has:*

(a) *Eigenvalues $\beta_0 = 1$, $\beta_j = \frac{n(n-1)\cdots(n-j+1)}{(\alpha+\beta+n)_j}, 1 \le j \le n$, $\beta_j = 0$ for $j > n$.*

(b) *Eigenfunctions $p_j$, the shifted Jacobi polynomials.*

(c) *With $z_i$ from* (4.3), *for any $\ell \ge 1$ and any starting state $\theta \in [0, 1]$,*

$$\chi_\theta^2(\ell) = \sum_{i=1}^{n} \beta_i^{2\ell} p_i^2(\theta) z_i.$$

The following proposition gives sharp chi-square bounds, uniformly over $\alpha, \beta, n$ in two cases: (i) $\alpha \ge \beta$, starting from 0 (worst starting point), (ii) $\alpha = \beta$, starting from $1/2$ (heuristically, the most favorable starting point). The restriction $\alpha \ge \beta$ is not really a restriction because of the symmetry $P_i^{a,b}(x) =$

$(-1)^i P_i^{b,a}(-x)$. For $\alpha \geq \beta > 1/2$, it is known (e.g., [63], Lemma 4.2.1) that

$$\sup_{[0,1]} |p_i| = \sup_{[-1,1]} |P_i^{\alpha-1,\beta-1}| = p_i(0) = \frac{(\alpha)_i}{i!}.$$

Hence, 0 is clearly the worst starting point from the viewpoint of convergence in chi-square distance, that is,

$$\sup_{\theta \in [0,1]} \{\chi_\theta^2(\ell)\} = \chi_0^2(\ell).$$

PROPOSITION 4.3. *For $\alpha \geq \beta > 0$, $n > 0$, set $N = \log[(\alpha + \beta)(\alpha + 1)/(\beta + 1)]$. The $\theta$-chain for the Beta/Binomial (4.2) satisfies:*

(i) 
- $\chi_0^2(\ell) \leq 7e^{-c}$, for $\ell \geq \frac{N+c}{-2\log \beta_1}$, $c > 0$.
- $\chi_0^2(\ell) \geq \frac{1}{6} e^c$, for $\ell \leq \frac{N-c}{-2\log \beta_1}$, $c > 0$.

(ii) *Assuming $\alpha = \beta > 0$,*
- $\chi_{1/2}^2(\ell) \leq 13\beta_2^{2\ell}$, for $\ell \geq \frac{1}{-2\log \beta_2}$.
- $\chi_{1/2}^2(\ell) \geq \frac{1}{2}\beta_2^{2\ell}$, for $\ell > 0$.

Roughly speaking, part (i) says that, starting from 0, $\ell(\alpha, \beta, n)$ steps are necessary and sufficient for convergence in chi-square distance where

$$\ell(\alpha, \beta, n) = \frac{\log[(\alpha + \beta)(\alpha + 1)/(\beta + 1)]}{-2\log(1 - (\alpha + \beta)/(\alpha + \beta + n))}.$$

Note that if $\alpha, n, n/\alpha$ tend to infinity and $\beta$ is fixed,

$$\ell(\alpha, \beta, n) \sim \frac{n \log \alpha}{\alpha}, \quad \beta_1 \sim 1 - \frac{\alpha}{n}.$$

If $\alpha, n, n/\alpha$ tend to infinity and $\alpha = \beta$,

$$\ell(\alpha, \alpha, n) \sim \frac{n \log \alpha}{4\alpha}, \quad \beta_1 \sim 1 - \frac{2\alpha}{n}.$$

The result also says that, starting from 0, convergence occurs abruptly (i.e., with cutoff) at $\ell(\alpha, \beta, n)$ as long as $\alpha$ tends to infinity.

Part (ii) indicates a completely different behavior starting from $1/2$ (in the case $\alpha = \beta$). There is no cutoff and convergence occurs at the exponential rate given by $\beta_2$ ($\beta_2 \sim 1 - \frac{4\alpha}{n}$ if $n/\alpha$ tends to infinity).

PROOF OF PROPOSITION 4.3(i). We have $\chi_0^2(\ell) = \sum_1^n \beta_i^{2\ell} p_i(0)^2 z_i$ and

$$\frac{\beta_{i+1}^{2\ell} p_{i+1}(0)^2 z_{i+1}}{\beta_i^{2\ell} p_i(0)^2 z_i}$$

$$= \left(\frac{n-i}{\alpha + \beta + n + i}\right)^{2\ell}$$

$$(4.4) \quad \cdot \frac{2i + \alpha + \beta + 1}{2i + \alpha + \beta - 1} \frac{i + \alpha + \beta - 1}{i + 1} \frac{i + \alpha}{i + \beta}$$

$$\leq \frac{5}{6} \frac{(\alpha + \beta)(\alpha + 1)}{\beta + 1}$$

$$\cdot \left(1 - \frac{\alpha + \beta + 2}{\alpha + \beta + n + 1}\right)^{2\ell}.$$

The lead term in $\chi_0^2(\ell)$ is

$$\left(\frac{(\alpha + \beta + 1)\alpha}{\beta}\right)\beta_1^{2\ell}.$$

From (4.4), we get that for any

$$\ell \geq \frac{1}{-2\log \beta_1} \log[(\alpha + \beta)(\alpha + 1)/(\beta + 1)]$$

we have

$$\frac{\beta_{i+1}^{2\ell} p_{i+1}(0)^2 z_{i+1}}{\beta_i^{2\ell} p_i(0)^2 z_i} \leq 5/6.$$

Hence, for such $\ell$,

$$\chi_0^2(\ell) \leq \left(\frac{(\alpha + \beta + 1)\alpha}{\beta}\right)\beta_1^{2\ell} \left(\sum_0^\infty (5/6)^k\right)$$

$$= 6\left(\frac{(\alpha + \beta + 1)\alpha}{\beta}\right)\beta_1^{2\ell}.$$

With $N = \log[(\alpha+\beta)(\alpha+1)/(\beta+1)]$ as in the proposition, we obtain

$$\chi_0^2(\ell) \leq 7e^{-c} \quad \text{for } \ell \geq \frac{N+c}{-2\log \beta_1}, \ c > 0;$$

$$\chi_0^2(\ell) \geq \tfrac{1}{6} e^c \quad \text{for } \ell \leq \frac{N-c}{-2\log \beta_1}, \ c > 0. \quad \square$$

PROOF OF PROPOSITION 4.3(ii). When $a = b$, the classical Jacobi polynomial $P_k^{a,b}$ is given by

$$P_k^{a,a}(x) = \frac{(a+1)_k}{(2a+1)_k} C_k^{a+1/2}(x)$$

where the $C_k^\nu$'s are the ultraspherical polynomials. See [63], (4.5.1). Now, [63], (4.5.16) gives $C_n^\nu(0) = 0$ if $n$ is odd and

$$C_n^\nu(0) = \frac{(2\nu)_n (-1)^{n/2}}{2^n (n/2)!(\nu + 1/2)_{n/2}}$$

if $n$ is even. Going back to the shifted Jacobi's, this yields $p_{2k+1}(1/2) = 0$ and

$$p_{2k}(1/2) = \frac{(\alpha)_{2k}}{(2\alpha - 1)_{2k}} C_{2k}^{\alpha - 1/2}(0)$$

$$= \frac{(\alpha)_{2k}}{(2\alpha - 1)_{2k}} \frac{(2\alpha - 1)_{2k}(-1)^k}{2^{2k} k!(\alpha)_k}$$

$$= \frac{(\alpha + k)_k (-1)^k}{2^{2k} k!}.$$

We want to estimate
$$\chi^2_{1/2}(\ell) = \sum_1^{\lfloor n/2 \rfloor} \beta_{2i}^{2\ell} p_{2i}(1/2)^2 z_{2i}$$

and thus we compute

$$\frac{\beta_{2(i+1)}^{2\ell} p_{2(i+1)}(1/2)^2 z_{2(i+1)}}{\beta_{2i}^{2\ell} p_{2i}(1/2)^2 z_{2i}}$$

$$= \left(\frac{(n-2i)(n-2i-1)}{(2\alpha+n+2i)(2\alpha+n+2i+1)}\right)^{2\ell}$$

$$\cdot \frac{4i+2\alpha+1}{4i+2\alpha-1} \frac{2i+2\alpha-1}{2i+2\alpha+1}$$

(4.5)
$$\cdot \frac{2i(2i+1)(2\alpha+2i+1)(2\alpha+2i)}{(2i+\alpha)^2(2i+\alpha+1)^2}$$

$$\cdot \left(\frac{(\alpha+2i)(\alpha+2i+1)}{4(\alpha+i)(i+1)}\right)^2$$

$$\leq \frac{9}{5}\beta_2^{2\ell}.$$

Hence
$$\chi^2_{1/2}(\ell) \leq 10\beta_2^{2\ell} p_2(1/2)^2 z_2 \quad \text{for } \ell \geq \frac{1}{-2\log\beta_2}.$$

As
$$p_2(1/2) = \frac{\alpha+1}{4} \quad \text{and} \quad z_2 = \frac{4(2\alpha+3)}{\alpha(\alpha+1)^2},$$

this gives $\chi^2_{1/2}(\ell) \geq \frac{1}{2}\beta_2^{2\ell}$ and, assuming $\ell \geq \frac{1}{-2\log\beta_2}$, $\chi^2_{1/2}(\ell) \leq 13\beta_2^{2\ell}$. □

### 4.2 Poisson/Gamma

4.2.1 *The x-chain for the Poisson/Gamma.* Fix $\alpha, a > 0$. For $x, y \in \mathcal{X} = \{0, 1, 2, \ldots\} = \mathbb{N}$, let

$$k(x,y) = \int_0^\infty \frac{e^{-\lambda(\alpha+1)/\alpha}\lambda^{a+x-1}}{\Gamma(a+x)(\alpha/(\alpha+1))^{a+x}}$$

(4.6)
$$\cdot \frac{e^{-\lambda}\lambda^y}{y!} d\lambda$$

$$= \frac{\Gamma(a+x+y)(\alpha/(2\alpha+1))^{a+x+y}}{\Gamma(a+x)(\alpha/(\alpha+1))^{a+x}y!}.$$

The stationary distribution is the negative binomial
$$m(x) = \frac{(a)_x}{x!}\left(\frac{1}{\alpha+1}\right)^a \left(\frac{\alpha}{\alpha+1}\right)^x, \quad x \in \mathbb{N}.$$

When $\alpha = a = 1$, the prior is a standard exponential, an example given in Section 2.1. Then,

$$k(x,y) = \left(\frac{1}{3}\right)^{x+y+1}\binom{x+y}{x}\bigg/\left(\frac{1}{2}\right)^{x+1},$$

$$m(x) = 1/2^{x+1}.$$

The orthogonal polynomials for the negative binomial are Meixner polynomials [68], (1.9):

$$M_j(x) = {}_2F_1\left(\begin{array}{c} -j, -x \\ a \end{array} \bigg| -\alpha\right).$$

These satisfy [68], (1.92),

$$\sum_{x=0}^\infty M_j(x) M_k(x) m(x) = \frac{(1+\alpha)^j j!}{(a)_j} \delta_{jk}.$$

Our work in previous sections, together with basic properties of Meixner polynomials, gives the following propositions.

PROPOSITION 4.4. *For $a, \alpha > 0$ the Poisson/Gamma x-chain* (4.6) *has:*

(a) *Eigenvalues $\beta_j = (\alpha/(1+\alpha))^j$, $0 \leq j < \infty$.*
(b) *Eigenfunctions $M_j(x)$, the Meixner polynomials.*
(c) *For any $\ell \geq 0$ and any starting state $x$*

$$\chi^2_x(\ell) = \sum_{y=0}^\infty \frac{(k^\ell(x,y) - m(y))^2}{m(y)}$$

$$= \sum_{i=1}^\infty \beta_i^{2\ell} M_i^2(x) z_i, \quad z_i = \frac{(a)_i}{(1+\alpha)^i i!}.$$

PROPOSITION 4.5. *For $\alpha = a = 1$, starting at $n$,*

$$\chi^2_n(\ell) \leq 2^{-2c} \quad \text{for } \ell = \log_2(1+n) + c, \ c > 0;$$
$$\chi^2_n(\ell) \geq 2^{2c} \quad \text{for } \ell = \log_2(n-1) - c, \ c > 0.$$

PROOF. From the definitions, for all $j$ and positive integer $x$,

$$|M_j(x)| = \left|\sum_{i=0}^{j \wedge x}(-1)^i \binom{j}{i} x(x-1)\cdots(x-i+1)\right|$$

$$\leq \sum_{i=0}^j \binom{j}{i} x^i = (1+x)^j.$$

Thus, for $\ell \geq \log_2(1+n) + c$,

$$\chi^2_n(\ell) = \sum_{j=1}^\infty M_j^2(n) 2^{-j(2\ell+1)}$$

$$\leq \sum_{j=1}^\infty (1+n)^{2j} 2^{-j(2\ell+1)}$$

$$\leq \frac{(1+n)^2 2^{-(2\ell+1)}}{1-(1+n)^2 2^{-(2\ell+1)}}$$

$$\leq \frac{2^{-2c-1}}{1-2^{-2c-1}} \leq 2^{-2c}.$$

The lower bound follows from using only the lead term. Namely,

$$\chi_n^2(\ell) \geq (1-n)^2 2^{-2\ell}$$
$$\geq 2^{2c} \quad \text{for } \ell = \log_2(n-1) - c. \quad \square$$

REMARK. Note the contrast with the Beta/Binomial example above. There, order $n$ steps are necessary and sufficient starting from $n$ and there is no cutoff. Here, $\log_2 n$ steps are necessary and sufficient and there is a cutoff. See [27] for further discussion of cutoffs.

Jim Hobert has shown us a neat appearance of the $x$-chain as branching process with immigration. Write the transition density above as

$$k(x,y) = \frac{\Gamma(a+x+y)}{\Gamma(a+x)y!}\left(\frac{\alpha+1}{2\alpha+1}\right)^{a+x}\left(\frac{\alpha}{2\alpha+1}\right)^y.$$

This is a negative binomial mass function with parameters $\theta = \frac{\alpha}{2\alpha+1}$ and $r = x + a$. Hence, the $x$-chain can be viewed as a branching process with immigration. Specifically, given that the population size at generation $n$ is $X_n = x$, we have

$$X_{n+1} = \sum_{i=1}^{x} N_{i,n} + M_{n+1},$$

where $N_{1,n}, N_{2,n}, \ldots, N_{x,n}$ are i.i.d. negative binomials with parameters $\theta = \frac{\alpha}{2\alpha+1}$ and $r = 1$, and $M_{n+1}$, which is independent of the $N_{i,n}$'s, has a negative binomial distribution with parameters $\theta = \frac{\alpha}{2\alpha+1}$ and $r = a$. The $N_{i,n}$'s represent the number of offspring of the $n$th generation and $M_{n+1}$ represents the immigration. This branching process representation was used in [61] to study admissibility.

4.2.2 *The $\theta$-chain for the Poisson/Gamma.* Fix $\alpha, a > 0$. For $\theta, \theta' \in \Theta = (0, \infty)$, let $\eta = (\alpha+1)\theta'/\alpha$. Note that $\pi(d\theta') = g(\theta')d\theta'$, where $g(\theta') = \frac{e^{-\theta'/\alpha}(\theta')^{a-1}}{\Gamma(a)\alpha^a}$. Hence, as in Section 4.1.2, we absorb $g(\theta')$ in the transition density of the $\theta$-chain. This gives

$$k(\theta, \theta') = \sum_{j=0}^{\infty} \frac{e^{-\theta}\theta^j}{j!} \frac{e^{-\theta'(\alpha+1)/\alpha}(\theta')^{a+j-1}}{\Gamma(a+j)(\alpha/(\alpha+1))^{a+j}}$$

$$= \frac{e^{-\theta-\theta'}(\theta')^{a-1}}{\alpha/(\alpha+1)} \sum_{j=0}^{\infty} \frac{(\theta\theta')^j}{j!\Gamma(a+j)}$$

$$= \frac{e^{-\theta-\theta'}}{\alpha/(\alpha+1)}\left(\frac{\theta'}{\theta}\right)^{(a-1)/2} \sum_{j=0}^{\infty} \frac{(\sqrt{\theta\theta'})^{2j+a-1}}{j!\Gamma(a+j)}$$

(4.7)

$$= \frac{e^{-\theta-\theta'}}{\alpha/(1+\alpha)}\left(\frac{\theta'}{\theta}\right)^{(a-1)/2} I_{a-1}(2\sqrt{\theta\theta'})$$

$$= \frac{e^{-\theta-(\alpha+1)\theta'/\alpha}}{\alpha/(1+\alpha)}\left(\frac{(\alpha+1)\theta'}{\alpha\theta}\right)^{(a-1)/2} I_{a-1}$$

$$\cdot (2\sqrt{(\alpha+1)\theta\theta'/\alpha}).$$

Thus, $k(\theta, \theta')$ is a transition density with respect to the Lebesgue measure $d\theta'$, that is,

$$\int k(\theta, \theta') d\theta' = 1.$$

Here $I_{a-1}$ is the modified Bessell function. For fixed $\theta$, $k(\theta, \theta')$ integrates to 1 as discussed in [44], pages 58–59. The stationary distribution of this Markov chain is the Gamma:

$$\pi(d\theta) = \frac{e^{-\theta/\alpha}\theta^{a-1}}{\Gamma(a)\alpha^a} d\theta.$$

To simplify notation, we take $\alpha = 1$ for the rest of this section. The relevant polynomials are the Laguerre polynomials ([68], Section 1.11)

$$L_i(\theta) = \frac{(a)_i}{i!} {}_1F_1\left(\begin{array}{c}-i\\a\end{array}\bigg|\theta\right)$$

$$= \frac{1}{i!}\sum_{j=0}^{i}\frac{(-i)_j}{j!}(a+j)_{i-j}\theta^j.$$

Note that classical notation has the parameter $a$ shifted by 1 whereas we have labeled things to mesh with standard statistical notation. The orthogonality relation is

$$\int_0^\infty L_i(\theta)L_j(\theta)\pi(\theta)d\theta = \frac{\Gamma(a+j)}{j!\Gamma(a)}\delta_{ij}$$
$$= z_j^{-1}\delta_{ij}.$$

The multilinear generating function formula ([63], Theorem 4.7.5) gives

$$\sum_{i=0}^{\infty} L_i(\theta)^2 z_i t^i = \frac{e^{-2t\theta/(1-t)}}{(1-t)^a}\sum_0^\infty \frac{1}{j!(a)_j}\left(\frac{\theta^2 t}{1-t^2}\right)^j.$$

Combining results, we obtain the following statements.

PROPOSITION 4.6. *For $\alpha = 1$ and $a > 0$, the Markov chain with kernel* (4.7) *has:*

(a) *Eigenvalues $\beta_j = \frac{1}{2^j}, 0 \leq j < \infty$.*
(b) *Eigenfunctions $L_j$, the Laguerre polynomials.*

(c) *For any $\ell \geq 1$ and any starting state $\theta$,*

$$\chi_\theta^2(\ell) = \sum_{j=1}^\infty \beta_j^{2\ell} L_j^2(\theta) \frac{j!\Gamma(a)}{\Gamma(a+j)}$$

$$= \frac{e^{-2^{-2\ell+1}\theta/(1-2^{-2\ell})}}{(1-2^{-2\ell})^a}$$

$$\cdot \sum_0^\infty \frac{1}{j!(a)_j} \left(\frac{\theta^2 2^{-2\ell}}{1-2^{-4\ell}}\right)^j - 1.$$

PROPOSITION 4.7. *For $\alpha = 1$ and $a > 0$, the Markov chain with kernel (4.7) satisfies:*

- *For $\theta > 0$, $\chi_\theta^2(\ell) \leq e^2 2^{-c}$ if $\ell \geq \frac{1}{2}(\log_2[2(1 + a + \theta^2/a)] + c)$, $c > 0$.*
- *For $\theta \in (0, a/2) \cup (2a, \infty)$, $\chi_\theta^2(\ell) \geq 2^c$ if $\ell \leq \frac{1}{2}(\log_2[\frac{1}{2}(\theta^2/a + a)] - c)$, $c > 0$.*

PROOF. For the upper bound, assuming $\ell \geq 1$, we write

$$\chi_\theta^2(\ell) = (1-4^{-\ell})^{-a} e^{-(2\theta 4^{-\ell})/(1-4^{-\ell})}$$

$$\cdot \sum_0^\infty \frac{1}{j!(a)_j} \left(\frac{\theta^2 4^{-\ell}}{1-4^{-\ell}}\right)^j - 1$$

$$\leq \frac{\exp((2\theta^2/a)4^{-\ell})}{(1-4^{-\ell})^a} - 1$$

$$\leq 2(\theta^2/a + a)4^{-\ell} \left(\frac{\exp(2(\theta^2/a)4^{-\ell})}{(1-4^{-\ell})^{a+1}}\right).$$

For $\ell \geq \frac{1}{2}(\log_2[2(1+\theta^2/a+a)]+c)$, $c > 0$, we obtain $\chi_\theta^2(\ell) \leq e^2 2^{-c}$.

The stated lower bound does not easily follow from the formula we just used for the upper bound. Instead, we simply use the first term in $\chi_\theta^2(\ell) = \sum_{j \geq 1} \beta_j^{2\ell} L_j^2(\theta) \frac{j!\Gamma(a)}{\Gamma(a+j)}$, that is, $a^{-1}(\theta-a)^2 4^{-\ell}$. This easily gives the desired result. □

REMARK. It is not easy to obtain sharp formulas starting from $\theta$ near $a$. For example, starting at $\theta = a$, one gets a lower bound by using the second term $\chi_\theta^2(\ell) = \sum_{j \geq 1} \beta_j^{2\ell} L_j^2(\theta) \frac{j!\Gamma(a)}{\Gamma(a+j)}$ (the first term vanishes at $\theta = a$). This gives $\chi_a^2(\ell) \geq [2a/(a+1)]4^{-2\ell}$. When $a$ is large, this is significantly smaller than the upper bound proved above.

### 4.3 The Gaussian Case

Here, the $x$-chain and the $\theta$-chain are essentially the same and indeed the same as the chain for the additive models, so we just treat the $x$-chain. Let $\mathcal{X} = \mathbb{R}$, $f_\theta(x) = e^{-1/2(x-\theta)^2/\sigma^2}/\sqrt{2\pi\sigma^2}$ and $\pi(d\theta) = \frac{e^{-1/2(\theta-\nu)^2/\tau^2}}{\sqrt{2\pi\tau^2}} d\theta$. The marginal density is Normal($\nu$, $\sigma^2 + \tau^2$).

A stochastic description of the chain is

$$X_{n+1} = aX_n + \varepsilon_{n+1}$$

(4.8)

$$\text{with } a = \frac{\tau^2}{\sigma^2+\tau^2}, \ \varepsilon \sim \text{Normal}\left(\frac{\sigma^2\nu}{\sigma^2+\tau^2}, \sigma^2\right).$$

This is the basic autoregressive (AR1) process. Feller ([44], pages 97–99) describes it as the discrete-time Ornstein–Uhlenbeck process. The diagonalization of this Gaussian Markov chain has been derived by other authors in various contexts. Goodman and Sokal [50] give an explicit diagonalization of vector-valued Gaussian autoregressive processes which specialize to (a), (b) below. Donoho and Johnstone ([31], Lemma 2.1) also specialize to (a), (b) below. Both sets of authors give further references. Since it is so well studied, we will be brief and treat the special case with $\nu = 0$, $\sigma^2 + \tau^2 = 1/2$. Thus the stationary distribution is Normal(0, 1/2). The orthogonal polynomials are now Hermite polynomials ([68], 1.13). These are given by

$$H_n(y) = (2y)^n {}_2F_0\left(\begin{array}{c}-n/2, -(n-1)/2\\--\end{array}\bigg| -\frac{1}{y^2}\right)$$

$$= n! \sum_{k=0}^{[n/2]} \frac{(-1)^k (2y)^{n-2k}}{k!(n-2k)!}.$$

They satisfy

$$\frac{1}{\sqrt{\pi}} \int_{-\infty}^\infty e^{-y^2} H_m(y) H_n(y) \, dy = 2^n n! \delta_{mn}.$$

There is also a multilinear generating function formula which gives ([63], Example 4.7.3)

$$\sum_0^\infty \frac{H_n(x)^2}{2^n n!} t^n = \frac{1}{\sqrt{1-t^2}} \exp\left(\frac{2x^2 t}{1+t}\right).$$

PROPOSITION 4.8. *For $\nu = 0$, $\sigma^2 + \tau^2 = 1/2$, the Markov chain (4.8) has:*

(a) *Eigenvalues $\beta_j = (2\tau^2)^j$ (as $\sigma^2 + \tau^2 = 1/2$, we have $2\tau^2 < 1$).*

(b) *Eigenfunctions the Hermite polynomials $H_j$.*

(c) *For any starting state $x$ and all $\ell \geq 1$,*

$$\chi_x^2(\ell) = \sum_{k=1}^\infty (2\tau^2)^{2k\ell} H_k^2(x) \frac{1}{2^k k!}$$

$$= \frac{\exp(2x^2(2\tau^2)^{2\ell}/(1+(2\tau^2)^{2\ell}))}{\sqrt{1-(2\tau^2)^{4\ell}}} - 1.$$

The next proposition turns the available chi-square formula into sharp estimates when $x$ is away from 0. Starting from 0, the formula gives $\chi_0^2(\ell) = (1-(2\tau^2)^{4\ell})^{-1/2}-1$. This shows convergence at the faster exponential rate of $\beta_2 = (2\tau^2)^2$ instead of $\beta_1 = 2\tau^2$.

PROPOSITION 4.9. *For $\nu=0$, $\sigma^2+\tau^2=1/2$, $x\in\mathbb{R}$, the Markov chain (4.8) satisfies:*

$$\chi_x^2(\ell) \leq 8e^{-c} \quad \text{for } \ell \geq \frac{\log(2(1+x^2))+c}{-2\log(2\tau^2)}, c>0,$$

$$\chi_x^2(\ell) \geq \frac{x^2 e^c}{2(1+x^2)} \quad \text{for } \ell \leq \frac{\log(2(1+x^2))-c}{-2\log(2\tau^2)},$$
$$c>0,$$

$$\chi_0^2(\ell) = (1-(2\tau^2)^{4\ell})^{-1}-1 \geq (2\tau^2)^{4\ell}.$$

PROOF. For the upper bound, assuming

$$\ell \geq \frac{1}{-2\log(2\tau^2)}(\log(2(1+x^2))+c), \quad c>0,$$

we have

$$(2\tau^2)^{2\ell} < 1/2, \quad 2x^2(2\tau^2)^{2\ell} < 1$$

and it follows that

$$\chi_x^2(\ell) = \frac{\exp(2x^2(2\tau^2)^{2\ell}/(1+(2\tau^2)^{2\ell}))}{\sqrt{1-(2\tau^2)^{4\ell}}} - 1$$
$$\leq (1+2(2\tau^2)^{4\ell})(1+6x^2(2\tau^2)^{2\ell}) - 1$$
$$\leq 8(1+x^2)(2\tau^2)^{2\ell}.$$

For the lower bound, write

$$\chi_x^2(\ell) = \frac{\exp(2x^2(2\tau^2)^{2\ell}/(1+(2\tau^2)^{2\ell}))}{\sqrt{1-(2\tau^2)^{4\ell}}} - 1$$
$$\geq \exp(x^2(2\tau^2)^{2\ell}) - 1$$
$$\geq x^2(2\tau^2)^{2\ell}. \qquad \square$$

## 5. LOCATION FAMILIES EXAMPLES

Sharp rates of convergence using spectral techniques are not restricted to exponential families and conjugate priors. In this section, we show how similar analyses are available for location families. Very different examples are collected in Section 6. In this section $f_\theta(x) = g(x-\theta)$ with $g$ and $\pi$ members of one of the six families of Section 2.4. To picture the associated Markov chains it is helpful to begin with the representation $X = \theta + \varepsilon$. Here $\theta$ is distributed as $\pi$ and $\varepsilon$ is distributed as $g$. The $x$-chain goes as follows: from $x$, draw $\theta'$ from $\pi(\cdot|x)$ and then go to $X' = \theta' + \varepsilon'$ with $\varepsilon'$ independently drawn from $g$. It has stationary distribution $m(x)\,dx$, the convolution of $\pi$ and $g$. For the $\theta$-chain, starting at $\theta$, set $X' = \theta + \varepsilon$ and draw $\theta'$ from $\pi(\cdot|x')$. It has stationary distribution $\pi$. Observe that

$$E_\theta(X^k) = E_\theta((\theta+\varepsilon)^k)$$
$$= \sum_{j=0}^{k} \binom{k}{j} \theta^j E(\varepsilon^{k-j}).$$

Thus (H2) of Section 3 is satisfied with $\eta_k = 1$. To check the conjugate condition we may use results of [87], Section 4. In present notation, Morris shows that if $p_k$ is the monic orthogonal polynomial of degree $k$ for the distribution $\pi$ and $p_k'$ is the monic orthogonal polynomial of degree $k$ for the distribution $m$, then

$$E_x(p_k(\theta)) = \left(\frac{n_1}{n_1+n_2}\right)^k b_k p_k'(x).$$

Here $\pi$ is taken as the sum of $n_1$ copies and $\varepsilon$ the sum of $n_2$ copies of one of the six families and

$$b_k = \prod_{i=0}^{k-1} \frac{1+ic/n_1}{1+ic/(n_1+n_2)},$$

where $c$ is the coefficient of $m^2(\theta)$ in $\sigma^2(\theta) = a + bm(\theta) + cm^2(\theta)$ for the family. Comparing lead terms gives (H3) (in Section 3) with an explicit value of $\mu_k$. In the present setup, $\mu_k = \beta_k$ is the $k$th eigenvalue.

We now make specific choices for each of the six cases.

### 5.1 Binomial

For fixed $p$, $0<p<1$, let $\pi = \text{Bin}(n_1,p)$, $g = \text{Bin}(n_2,p)$. Then $m = \text{Bin}(n_1+n_2,p)$ and

$$\pi(\theta|x) = \frac{\binom{n_1}{\theta}\binom{n_2}{x-\theta}}{\binom{n_1+n_2}{x}}$$

is hypergeometric. The $\theta$-chain progresses as a population process on $0\leq\theta\leq n_1$: from $\theta$, there are $\varepsilon$ new births and the resulting population of size $x = \theta+\varepsilon$ is thinned down by random sampling. The $x$-chain denoted by $\{X_k\}_{k\geq 0}$ can be represented in an autoregressive cast. More precisely,

(5.1) $$X_{k+1} = S_{X_k} + \varepsilon_{k+1},$$

where $S_{X_k}$ is a hypergeometric with parameters $n_1, n_2, X_k$ and $\varepsilon_{k+1}$ is drawn independently from $\text{Bin}(n_2,p)$.

For the binomial, the parameter $c$ is $c = -1$ and the eigenvalues of the $x$-chain are

$$\beta_k = \frac{n_1(n_1-1)\cdots(n_1-k+1)}{(n_1+n_2)(n_1+n_2-1)\cdots(n_1+n_2-k+1)},$$

$$0 \leq k \leq n_1 + n_2 = N.$$

Note that $\beta_k = 0$ for $k \geq n_1 + 1$. The orthogonal polynomials are Krawtchouck polynomials ([68], 1.10; [63], page 100):

(5.2) $$k_j(x) = {}_2F_1\left(\begin{matrix}-j, -x \\ -N\end{matrix}\,\bigg|\,\frac{1}{p}\right)$$

which satisfy

$$\sum_{x=0}^{N} \binom{N}{x} p^x (1-p)^{N-x} k_j(x) k_\ell(x)$$

$$= \binom{N}{j}^{-1}\left(\frac{1-p}{p}\right)^j \delta_{j\ell}.$$

PROPOSITION 5.1. *Consider the chain* (5.1) *on* $\{0,\ldots,n_1+n_2\}$ *with* $0 < p < 1$, *starting at* $x = 0$. *Set* $N = n_1 + n_2$, $q = p/(1-p)$. *Then we have*

$$e^{-c} \leq \chi_0^2(\ell) \leq e^{-c} e^{e^{-c}}$$

*whenever*

$$\ell = \frac{\log(qN) + c}{-2\log(1 - n_2/N)}, \quad c \in (-\infty, \infty).$$

Note two cases of interest: (i) For $p = 1/2$, the proposition shows that $\frac{\log(N)}{-2\log(1-n_2/N)}$ steps are necessary and sufficient. There is a chi-square cutoff when $N$ tends to infinity. (ii) For $p = 1/N$, there is no cutoff.

PROOF. From (2.9) and (5.2), we have $k_j^2(0) = 1$ for all $j$ and the chi-square distance becomes

$$\chi_0^2(\ell) = \sum_{j=1}^{N} \beta_j^{2\ell}\binom{N}{j} q^j$$

with $N = n_1 + n_2$, $q = p/(1-p)$ and $\ell \geq 1$. For $j \leq n_1$, the eigenvalues satisfy

$$\beta_j = \prod_{i=0}^{j-1}\left(1 - \frac{n_2}{N-i}\right) \leq \left(1 - \frac{n_2}{N}\right)^j = \beta_1^j.$$

Hence, we obtain

$$\chi_0^2(\ell) \leq \sum_{j=1}^{N}\binom{N}{j}(q\beta_1^{2\ell})^j$$

$$= (1 + q\beta_1^{2\ell})^N - 1$$

$$\leq qN\beta_1^{2\ell}(1 + q\beta_1^{2\ell})^{N-1}.$$

This gives the desired result since we also have $\chi_0^2(\ell) \geq Nq\beta_1^{2\ell}$.  □

In general, it is not very easy to evaluate the polynomials $k_j$ at $x \neq 0$ to estimate $\chi_x^2(\ell)$ and understand what the role of the starting point is. However, if $p = 1/2$ and $x = n_1 = N/2$, then the recurrence equation ([63], (6.2.37)) shows that $k_{2j+1}(N/2) = 0$ and

$$k_{2j}(N/2) = (-1)^j\frac{(2j-1)!}{(N-2j+1)_{2j}}.$$

Thus, for $\ell \geq 1$,

$$\chi_{N/2}^2(\ell) = \sum_{j=1}^{\lceil N/4 \rceil} \beta_{2j}^{2\ell}\binom{N}{2j}\left(\frac{(2j-1)!}{(N-2j+1)_{2j}}\right)^2$$

$$= \sum_{j=1}^{\lceil N/4 \rceil} \beta_{2j}^{2\ell} \frac{(2j-1)!}{2j(N-2j+1)_{2j}}.$$

This is small for $\ell = 1$ if $N$ is large enough. Indeed, splitting the sum into two parts at $N/8$ easily yields

$$\chi_{N/2}^2(\ell) \leq \frac{3}{2N}2^{-4\ell} + N2^{-N\ell/2}.$$

**5.2 Poisson**

Fix positive reals $\mu, n_1, n_2$. Let $\pi = \text{Poisson}(\mu n_1)$, $g = \text{Poisson}(\mu n_2)$. Then

$$m = \text{Poisson}(\mu(n_1 + n_2))$$

and

$$\pi(\theta|x) = \text{Bin}\left(x, \frac{n_1}{n_1 + n_2}\right).$$

The $x$-chain is related to the $M/M/\infty$ queue and the $\theta$-chain is related to Bayesian missing data examples in Section 2.3.3. Here, the parameter $c = 0$ so that

$$\beta_k = \left(\frac{n_1}{n_1+n_2}\right)^k, \quad 0 \leq k < \infty.$$

The orthogonal polynomials are Charlier polynomials ([68], 1.12; [63], page 177):

$$C_j(x) = {}_2F_0\left(\begin{matrix}-j, -x \\ -\end{matrix}\,\bigg|\,-\frac{1}{\mu}\right),$$

$$\sum \frac{e^{-\mu}\mu^x}{x!}C_j(x)C_k(x) = j!\mu^{-j}\delta_{jk}.$$

We carry out a probabilistic analysis of this problem in [25].

### 5.3 Negative Binomial

Fix $p$ with $0 < p < 1$ and positive real $n_1, n_2$. Let $\pi = \text{NB}(n_1, p)$, $g = \text{NB}(n_2, p)$. Then $m = \text{NB}(n_1 + n_2, p)$ and

$$\pi(\theta|x) = \binom{x}{\theta} \frac{\Gamma(n_1+n_2)\Gamma(\theta+n_1)\Gamma(x-\theta-n_2)}{\Gamma(x+n_1+n_2)\Gamma(n_1)\Gamma(n_2)},$$
$$0 \le \theta \le x,$$

which is a negative hypergeometric. A simple example has $n_1 = n_2 = 1$ (geometric distribution) so $\pi(\theta|x) = 1/(1+x)$. The $x$-chain becomes: From $x$, choose $\theta$ uniformly in $0 \le \theta \le x$ and let $X' = \theta + \varepsilon$ with $\varepsilon$ geometric. The parameter $c = 1$ so that

$$\beta_0 = 1,$$
$$\beta_k = \frac{n_1(n_1+1)\cdots(n_1+k-1)}{(n_1+n_2)(n_1+n_2+1)\cdots(n_1+n_2+k-1)},$$
$$1 \le k < \infty.$$

The orthogonal polynomials are Meixner polynomials discussed in Section 4.2.

### 5.4 Normal

Fix reals $\mu$ and $n_1, n_2, v > 0$. Let $\pi = \text{Normal}(n_1\mu, n_1 v)$, $g = \text{Normal}(n_2\mu, n_2 v)$. Then $m = \text{Normal}((n_1+n_2)\mu, (n_1+n_2)v)$ and $\pi(\theta|x) = \text{Normal}(\frac{n_1}{n_1+n_2}x, \frac{n_1 n_2}{n_1+n_2}v)$. Here $c = 0$ and

$$\beta_k = \left(\frac{n_1}{n_1+n_2}\right)^k, \quad 0 \le k < \infty.$$

The orthogonal polynomials are Hermite, discussed in Section 4.3. Both the $x$- and $\theta$-chains are classical autoregressive processes as described in Section 4.3.

### 5.5 Gamma

Fix positive real $n_1, n_2, \alpha$. Let $\pi = \text{Gamma}(n_1, \alpha)$, $g = \text{Gamma}(n_2, \alpha)$. Then

$$m = \text{Gamma}(n_1+n_2, \alpha),$$
$$\pi(\theta|x) = x \cdot \text{Beta}(n_1, n_2).$$

A simple case to picture is $\alpha = n_1 = n_2 = 1$. Then, the $x$-chain may be described as follows: From $x$, choose $\theta$ uniformly in $(0, x)$ and set $X' = \theta + \varepsilon$ with $\varepsilon$ standard exponential. This is simply a continuous version of the examples of Section 5.3. The parameter $c = 1$ and so

$$\beta_0 = 1,$$
$$\beta_k = \frac{n_1(n_1+1)\cdots(n_1+k-1)}{(n_1+n_2)(n_1+n_2+1)\cdots(n_1+n_2+k-1)},$$
$$0 < k < \infty.$$

The orthogonal polynomials are Laguerre polynomials, discussed in Section 4.2 above.

### 5.6 Hyperbolic

The density of the sixth family is given in Section 2.3 in terms of parameters $r > 0$ and $|\theta| < \pi/2$. It has mean $\mu = r\tan(\theta)$ and variance $\mu^2/r + r$. See [86], Section 5 or [37] for numerous facts and references. Fix real $\mu$ and positive $n_1, n_2$. Let the density $\pi$ be hyperbolic with mean $n_1\mu$ and $r_1 = n_1(1+\mu^2)$. Let the density $g$ be hyperbolic with mean $n_2\mu$ and $r_2 = n_2(1+\mu^2)$. Then $m$ is hyperbolic with mean $(n_1+n_2)\mu$ and $r = (n_1+n_2)(1+\mu^2)$. The conditional density $\pi(\theta|x)$ is "unnamed and apparently has not been studied" ([87], page 581).

For this family, the parameter $c = 1$ and thus

$$\beta_0 = 1,$$
$$\beta_k = \frac{n_1(n_1+1)\cdots(n_1+k-1)}{(n_1+n_2)\cdots(n_1+n_2+k-1)}.$$

The orthogonal polynomials are Meixner–Pollaczek polynomials ([100], page 395; [68], 1.7; [63], page 171). These are given in the form

$$P_n^\lambda(x, \varphi)$$
$$= \frac{(2\lambda)_n}{n!} {}_2F_1\left(\begin{array}{c}-n, \lambda+ix \\ 2\lambda\end{array}\bigg| 1-e^{-2i\varphi}\right) e^{in\varphi},$$

(5.3)
$$\frac{1}{2\pi}\int_{-\infty}^{\infty} e^{(2\varphi-\pi)x}|\Gamma(\lambda+ix)|^2 P_m^\lambda P_n^\lambda \, dx$$
$$= \frac{\Gamma(n+2\lambda)}{n!(2\sin\varphi)^{2\lambda}}\delta_{mn}.$$

Here $-\infty < x < \infty$, $\lambda > 0$, $0 < \varphi < \pi$. The change of variables $y = \frac{rx}{2}$, $\varphi = \frac{\pi}{2} + \tan^{-1}(\theta)$ $\lambda = r/2$ transforms the density $e^{(2\varphi-\pi)x}|\Gamma(\lambda+ix)|^2$ to a constant multiple of the density $f_\theta(x)$ of Section 2.4.

We carry out one simple calculation. Let $\pi, g$ have the density of $\frac{2}{\pi}\log|C|$, with $C$ standard Cauchy. Thus

(5.4) $$\pi(dx) = g(x)\,dx = \frac{1}{2\cosh(\pi x/2)}\,dx.$$

The marginal density is the density of $\frac{2}{\pi}\log|C_1 C_2|$, that is,

$$m(x) = \frac{x}{2\sinh(\pi x/2)}.$$

PROPOSITION 5.2. *For the additive walk based on* (5.4):

(a) *The eigenvalues are* $\beta_k = \frac{1}{k+1}$, $0 \le k < \infty$.

(b) *The eigenfunctions are the Meixner–Pollaczek polynomials* (5.3) *with* $\varphi = \pi/2$, $\lambda = 1$.

(c) $\chi_x^2(\ell) = 2\sum_{k=1}^{\infty}(k+1)^{-2\ell-1}(P_k^1(\frac{x}{2},\frac{\pi}{2}))^2$.

PROOF. Using $\Gamma(z+1) = z\Gamma(z)$, $\Gamma(z)\Gamma(1-z) = \frac{\pi}{\sin(\pi z)}$, we check that

$$|\Gamma(1+ix)|^2 = \Gamma(1+ix)\Gamma(1-ix)$$
$$= (ix)\Gamma(ix)\Gamma(1-ix)$$
$$= \frac{\pi(ix)}{\sin\pi(ix)} = \frac{\pi x}{\sinh(\pi x)}.$$

The result now follows from routine simplification. □

REMARK. Part (c) has been used to show that order $\log x$ steps are necessary and sufficient for convergence in unpublished joint work with Mourad Ismail.

## 6. OTHER MODELS, OTHER METHODS

Even in the limited context of bivariate Gibbs sampling, there are other contexts in which the ingredients above arise. These give many further examples where present techniques lead to sharp results. This section gives brief pointers to Lancaster families, alternating projections, the multivariate case and to other techniques for proving convergence.

### 6.1 Lancaster Families

There has been a healthy development of bivariate distributions with given margins. The part of interest here begins with the work of Lancaster, nicely summarized in his book [72]. Relevant papers by Angelo Koudou [69, 70, 71] summarize recent work. Briefly, let $(\mathcal{X},\mu), (\mathcal{Y},\nu)$ be probability spaces with associated $L^2(\mu), L^2(\nu)$. Let $\sigma$ be a measure on $\mathcal{X} \times \mathcal{Y}$ with margins $\mu$ and $\nu$. Suppose the relevant conditional probabilities exist, so

$$\sigma(dx,dy) = \mu(dx)K_x(dy) = \nu(dy)L_y(dx).$$

Say that $\sigma$ is a Lancaster probability with respect to the given sequences of orthonormal functions if there exists a positive real sequence $\{\rho_n\}$ such that, for all $n$,

$$(6.5) \qquad \int p_n(x)q_m(y)\sigma(dx,dy) = \rho_n \delta_{mn}.$$

This equation is also equivalent to $\int q_n(y)K_x(dy) = \rho_n p_n(x)$ and to $\int p_n(x)K_y(dx) = \rho_n q_n(y)$.

Koudou shows that, given $\sigma$, one can always find sequences of orthonormal functions such that $\sigma$ is Lancaster for these sequences. He then characterizes those sequences $\rho_n$ such that the associated $\sigma$ is absolutely continuous with respect to $\mu \times \nu$ with

$$(6.6) \qquad f = \frac{d\sigma}{d(\mu \times \nu)} = \sum_n \rho_n p_n(x)q_n(y)$$

in $L^2(\mu \times \nu)$.

For such Lancaster densities (6.6), the equivalence (6.5) says precisely that the $x$-chain for the Gibbs sampler for $f$ has $\{p_n\}$ as eigenvectors with eigenvalues $\{\rho_n^2\}$ (cf. Section 3 above). For more on Lancaster families with marginals in the six exponential families, see [7], Section 7, which makes fascinating connections between these families and diagonal multivariate families.

The above does not require polynomial eigenvectors and Griffiths [52] gives examples of triangular margins and explicit nonpolynomial eigenfunctions. Here is another. Let $G$ be a compact Abelian group with characters $\{p_n(x)\}$ chosen orthonormal with respect to Haar measure $\mu$. For a probability density $f$ on $G$, the location problem

$$\sigma(dx,dy) = f(x-y)\mu(dx)\nu(dy)$$

has uniform margins, $\{p_n\}$ as eigenfunctions of the $x$-chain and $\rho_n = \int p_n(x)f(x)\,d\mu(x)$ as eigenvalues.

A main focus of Koudou and his co-authors is delineating all of the extremal Lancaster sequences and so, by Choquet's theorem (every point in a compact convex subset of a metrizable topological vector space is a barycenter of a probability measure supported on the extreme points; see [84], Chapter 11.2) all densities $f$ with $\{p_n\}, \{q_n\}$ as in (6.5). Of course, any of these can be used with the Gibbs sampler and present techniques. These characterizations are carried out for a variety of classical orthogonal polynomials. In particular, Koudou gives a clear translation into probabilists' language of Gasper's complete determination of the extremals of the Lancaster families with equal Beta$(\alpha,\beta)$ margins for $\alpha,\beta \geq \frac{1}{2}$. We give but one example.

EXAMPLE. Consider the uniform distribution on the unit disk in $\mathbb{R}^2$ given by

$$(6.7) \qquad f(x,y) = \frac{1}{\pi}, \quad 0 \leq x^2 + y^2 \leq 1.$$

The conditional distribution of $X$ given $Y$ is uniform on $[-\sqrt{1-Y^2}, \sqrt{1-Y^2}]$. Similarly, the conditional distribution of $Y$ given $X$ is uniform on

$[-\sqrt{1-X^2}, \sqrt{1-X^2}]$. The marginal density of $X$ is

$$m(x) = \frac{2}{\pi}\sqrt{1-x^2}, \quad -1 \leq x \leq 1.$$

By symmetry, $Y$ has the same marginal density. Since for every $k \geq 0$,

$$\mathbf{E}(X^{2k} \mid Y) = \frac{(1-Y^2)^k}{2k+1},$$

$$\mathbf{E}(Y^{2k} \mid X) = \frac{(1-X^2)^k}{2k+1}$$

and

$$\mathbf{E}(X^{2k+1} \mid Y) = 0,$$
$$\mathbf{E}(Y^{2k+1} \mid X) = 0,$$

we conclude that the $x$-chain has polynomial eigenfunctions $\{p_k(x)\}_{k\geq 0}$ and eigenvalues $\{\lambda_k\}_{k\geq 0}$ where $\lambda_{2k} = \frac{1}{(2k+1)^2}$ and $\lambda_{2k+1} = 0, k \geq 0$. The polynomials $\{p_k\}_{k\geq 0}$ are orthogonal polynomials corresponding to the marginal density $m$. They are Chebyshev polynomials of the second kind, given by the identity

$$p_k(\cos(\theta)) = \frac{\sin((k+1)\theta)}{\sin(\theta)}, \quad 0 \leq \theta \leq \pi, k \geq 0.$$

Using present theory we have the following theorem.

THEOREM 6.1. *For the $x$-chain from the Gibbs sampler for the density* (6.7):

(a) $\chi_x^2(l) = \sum_{k=1}^{\infty} \frac{1}{(2k+1)^{4l}} p_{2k}^2(x)$.
(b) *For* $x = 0$, $\chi_x^2(l) = \sum_{k=1}^{\infty} \frac{1}{(2k+1)^{4l}}$.
*Thus*

$$\frac{1}{3^{4l}} \leq \chi_x^2(l) \leq \left(\frac{8l+1}{8l-2}\right) \frac{1}{3^{4l}}.$$

(c) *For* $|x| = 1$, $\chi_x^2(l) = \sum_{k=1}^{\infty} \frac{1}{(2k+1)^{4l-2}}$.
*Thus*

$$\frac{1}{3^{4l-2}} \leq \chi_x^2(l) \leq \left(\frac{8l-3}{8l-6}\right) \frac{1}{3^{4l-2}}.$$

One of the hallmarks of our examples is that they are statistically natural. It is an open problem to give natural probabilistic interpretations of a general Lancaster family. There are some natural examples. Eagleson [36] has shown that $W_1 + W_2$ and $W_2 + W_3$ are Lancaster if $W_1, W_2, W_3$ are independent and from one of the six quadratic families. Natural Markov chains with polynomial eigenfunctions have been extensively studied in mathematical genetics literature. This work, which perhaps begins with [42], was unified in [14]. See [38] for a textbook treatment. Models of Fisher–Wright, Moran, Kimura, Karlin and McGregor are included. While many models are either absorbing, nonreversible, or have intractable stationary distributions, there are also tractable new models to be found. See the Stanford thesis work of Hua Zhou.

Interesting classes of reversible Markov chains with explicit polynomial eigenfunctions appear in the work of Hoare, Rahman and their collaborators [20, 58, 59, 60]. These seem distinct from the present examples, are grounded in physics problems (transfer of vibrational energy between polyatomic molecules) and involve some quite exotic eigenfunctions ($9 - j$ symbols). Their results are explict and it seems like a worthwhile project to convert them into sharp rates of convergence.

A rather different class of examples can be created using autoregressive processes. For definiteness, work on the real line $\mathbb{R}$. Consider processes of form $X_0 = 0$, and for $1 \leq n < \infty$,

$$X_{n+1} = a_{n+1}X_n + \varepsilon_{n+1},$$

with $\{(a_i, \varepsilon_i)\}_{i\geq 1}$ independent and identically distributed. Under mild conditions on the distribution of $(a_i, \varepsilon_i)$, the Markov chain $X_n$ has a unique stationary distribution $m$ which can be represented as the probability distribution of

$$X_\infty = \varepsilon_0 + a_0 \varepsilon_1 + a_1 a_0 \varepsilon_2 + \cdots.$$

The point here is that for any $k$ such that moments exist

$$E(X_1^k | X_0 = x) = E((a_1 x + \varepsilon_1)^k)$$
$$= \sum_{i=0}^{k} \binom{k}{i} x^i E(a_1^i \varepsilon_1^{k-i}).$$

If, for example, the stationary distribution $m$ has moments of all orders and is determined by those moments, then the Markov chain $\{X_n\}_{n=0}^{\infty}$ is generated by a compact operator with eigenvalues $E(a_1^i)$, $0 \leq i < \infty$, and polynomial eigenfunctions.

We have treated the Gaussian case in Section 4.5. At the other extreme, take $|a| < 1$ constant and let $\varepsilon_i$ take values $\pm 1$ with probability $1/2$. The fine properties of $\pi$ have been intensively studied as Bernoulli convolutions. See [23] and the references there. For example, if $a = 1/2$, then $\pi$ is the usual uniform distribution on $[-1, 1]$ and the polynomials are Chebyshev polynomials. Unfortunately, for any value of $a \neq 0$, in the $\pm 1$ case, the distribution $\pi$ is known to be continuous while the distribution of $X_n$ is discrete and so does not converge to $\pi$ in $L^1$ or $L^2$. We do not know how to use the eigenvalues to get quantitative rates of convergence in one of the standard metrics for weak convergence.

As a second example take $(a, \varepsilon) = (u, 0)$ with probability $p$ and $(1 + u, -u)$ with probability $1 - p$ with $u$ uniform on $(0, 1)$ and $p$ fixed in $(0, 1)$. This Markov chain has a Beta$(p, 1 - p)$ stationary density. The eigenvalues are $1/(k + 1)$, $1 \leq k < \infty$. It has polynomial eigenfunctions. Alas, it is not reversible and again we do not know how to use the spectral information to get usual rates of convergence. See [23] or [75] for more information about this so-called "donkey chain."

Finally, we mention the work of Hassairi and Zarai [57] which develops the orthogonal polynomials for cubic (and other) exponential families such as the inverse Gaussian. They introduce a novel notion of 2-orthogonality. It seems possible (and interesting) to use their tools to handle the Gibbs sampler for conjugate priors for cubic families.

### 6.2 Alternating Conditional Expectations

The alternating conditional expectations that underlie the Gibbs sampler arise in other parts of probability and statistics. These include classical canonical correlations, especially those abstracted by Dauxois and Pousse [21]. This last paper contains a framework for studying our problems where the associated operators are not compact.

A nonlinear version of canonical correlations was developed by Breiman and Jerry Friedman as the A.C.E. algorithm. Buja [13] pointed out the connections to Lancaster families. He found several other parallel developments, particularly in electrical engineering ([13], Section 7–12). Many of these come with novel, explicit examples which are grist for our mill. Conversely, our development gives new examples for understanding the alternating conditional expectations that are the central focus of [13].

There is a classical subject built around alternating projections and the work of von Neumann. See Deutsch [22]. Let $\mathcal{H}_1$ and $\mathcal{H}_2$ be closed subspaces of a Hilbert space $\mathcal{H}$. Let $P_1$, $P_2$ be the orthogonal projection onto $\mathcal{H}_1$, $\mathcal{H}_2$ and let $P_I$ be the orthogonal projection onto $\mathcal{H}_1 \cap \mathcal{H}_2$. von Neumann showed that $(P_1 P_2)^n \to P_I$ as $n$ tends to infinity. That is $\|(P_1 P_2)^n(x) - P_I(x)\| \to 0$. In [24] we show that the Gibbs sampler is a special case of von Neumann's algorithm; with $\mathcal{H} = L^2(\sigma), \mathcal{H}_1 = L^2(\mu), \mathcal{H}_2 = L^2(\nu)$ using the notation of Section 6.1. We develop the tools used to quantify rates of convergence for von Neumann's algorithm for use with the Gibbs sampler in [24].

### 6.3 Multivariate Models

The present paper and its companion paper [25] have discussed univariate models. There are a number of models with $x$ or $\theta$ multivariate where the associated Markov chains have polynomial eigenfunctions. Some analogs of the six exponential families are developed in [15]. In Koudou and Pommeret [69], the Lancaster theory for these families is elegantly developed. Their work can be used to give the eigenvalues and eigenfunctions for the multivariate versions of the location models in Section 5. In their work, families of multivariate polynomials depending on a parameter matrix are given. For our examples, specific forms must be chosen. In a series of papers, developed independently of [69], Griffiths [51, 53] has developed such specific bases and illuminated their properties. This work is turned into sharp rates of convergence for two-component multivariate Gibbs samplers in the Stanford thesis work of Kshitij Khare and Hua Zhou.

An important special case, high-dimensional Gaussian distributions, has been studied in [2, 50]. Here is a brief synopsis of these works. Let $m(x)$ be a $p$-dimensional normal density with mean $\mu$ and covariance $\Sigma$ [i.e., $N_p(\mu, \Sigma)$]. A Markov chain with stationary density $m$ may be written as

$$(6.8) \qquad X_{n+1} = AX_n + Bv + C\varepsilon_{n+1}.$$

Here $\varepsilon_n$ has a $N_p(0, I)$ distribution, $v = \Sigma^{-1}\mu$, and the matrices $A, B, C$ have the form

$$A = -(D + L)^{-1}L^T,$$
$$B = (D + L)^{-1},$$
$$C = (D + L)^{-1}D^{1/2},$$

where $D$ and $L$ are the diagonal and lower triangular parts of $\Sigma^{-1}$. The chain (6.8) is reversible if and only if $A\Sigma = \Sigma A^T$. If this holds, $A$ has real eigenvalues $(\lambda_1, \lambda_2, \ldots, \lambda_p)$. In [50], Goodman and Sokal show that the Markov chain (6.8) has eigenvalues $\lambda^K$ and eigenfunctions $H_K$ for $K = (k_1, k_2, \ldots, k_p)$, $k_i \geq 0$, with

$$\lambda^K = \prod_{i=1}^{p} \lambda_i^{k_i}, \quad H_K(x) = \prod_{i=1}^{p} H_{k_i}(z_i),$$

where $Z = P^T \Sigma^{-1/2} X$ and $\{H_k\}$ are the usual one-dimensional Hermite polynomials. Here $\Sigma^{-1/2} A \Sigma^{1/2} = PDP^T$ is the eigendecomposition of $\Sigma^{-1/2} A \Sigma^{1/2}$. Goodman and Sokal show how a variety of stochastic algorithms, including the systematic scan Gibbs sampler for sampling from $m$, are covered by this framework. Explicit rates of convergence for this Markov chain can be found in the Stanford thesis work of Kshitij Khare.

## 6.4 Conclusion

The present paper studies rates of convergence using spectral theory. In a companion paper [25] we develop a stochastic approach which uses one eigenfunction combined with coupling. This is possible when the Markov chains are stochastically monotone. We show this is the case for all exponential families, with any choice of prior, and for location families where the density $g(x)$ is totally positive of order 2. This lets us give rates of convergence for the examples of Section 4 when moments do not exist (negative binomial, gamma, hyperbolic). In addition, location problems fall into the setting of iterated random functions so that backward iteration and coupling are available. See [17, 23] for extensive references.

## APPENDIX: PROOF OF THEOREM 3.1

The proof will follow from the two dual lemmas below which show that the expectation operators $E_\theta$ and $E_x$ each take one orthogonal polynomial family into the other. For the Beta/Binomial example treated in the Introduction and in Section 4.1, these operators relate Hahn polynomials on $\{0, \ldots, n\}$ to Jacobi polynomials on $(0, 1)$. These facts are of independent interest and some have been observed before. See [62], (3.7) for the correspondence between Hahn and Jacobi polynomials and for a host of further references.

LEMMA A1. $E_\theta[p_k(X)] = \eta_k q_k(\theta)$, $0 \leq k < c$.

PROOF. For $k = 0$, $E_\theta[p_0] = 1 = \eta_0 q_0$. If $0 < k < c$, then for $0 \leq i < k$, the unconditional expectation is given by

$$E[\theta^i p_k(X)] = E[p_k(X) E_X(\theta^i)] = E[p_k(X) \widehat{p}(X)]$$

with $\widehat{p}$ a polynomial of degree $i < k$. Since $0 \leq i < k < c$, $E[p_k(X) \widehat{p}(X)] = 0$ by orthogonality. Thus $0 = E[\theta^i p_k(X)] = E[\theta^i E_\theta(p_k(X))]$. By assumption (H2), $\eta_k^{-1} E_\theta[p_k(X)]$ is a monic polynomial of degree $k$ in $\theta$. Since it is orthogonal to all polynomials of degree less than $k$, we must have $E_\theta[p_k(X)] = \eta_k q_k(\theta)$. □

The second lemma is dual to the first.

LEMMA A2. $E_x[q_k(\theta)] = \mu_k p_k(x)$, $0 \leq k < c$. If $c < \infty$, $E_x(q_k(\theta)) = 0$ for $k \geq c$.

PROOF. The first part is proved as per Lemma A1. If $c < \infty$, and $k \geq c$, by the same argument we have, for $0 \leq j < c$, $E[p_j(X) E_X[q_k(\theta)]] = 0$. But $\{p_j\}_{0 \leq j < c}$ form a basis for $L^2(m(dx))$, and $E_x[q_k(\theta)] \in L^2(m(dx))$ since

$$E[(E_X q_k(\theta))^2] \leq E[q_k^2(\theta)] < \infty.$$

It follows that $E_x[q_k(\theta)] = 0$. □

PROOF OF PART (a) OF THEOREM 3.1. Suppose $0 \leq k < c$. From the definitions, the $x$-chain operates on $p_k$ as

$$E_x[E_\theta(p_k(X'))] = E_x[\eta_k q_k(\theta)] = \eta_k \mu_k p_k(x)$$

with equalities from Lemmas A1, A2. Hence, $\eta_k \mu_k$ are eigenvalues of the $x$-chain with $p_k$ as eigenfunctions. This proves (a). □

PROOF OF PART (b). Suppose first $0 \leq k < c$. Then, arguing as above, $\mu_k \eta_k$ are eigenvalues of the $\theta$-chain with $q_k$ as eigenvectors. If $c = \infty$, we are done. If $c < \infty$, then, for $k \geq c$, Lemma A2 shows that $q_k$ is an eigenfunction for the $\theta$-chain with eigenvalue zero. □

PROOF OF PART (c). From the development in Section 2.1, the random scan chain $K$ takes $L^2(P)$ into $L^2(m) + L^2(\pi) \subseteq L^2(P)$ and ker K $\supseteq (L^2(m) + L^2(\pi))^\perp$. We have

$$Kg(X, \theta) = \tfrac{1}{2} E_x[g(x, \theta')] + \tfrac{1}{2} E_\theta[g(X', \theta)].$$

For $0 \leq k < c$, consider $K$ acting on $p_k(x) + \sqrt{\frac{\eta_k}{\mu_k}} q_k(\theta)$. The result is

$$\frac{1}{2}\left(p_k(x) + E_x[q_k(\theta')]\sqrt{\frac{\eta_k}{\mu_k}}\right)$$
$$+ \frac{1}{2}\left(E_\theta[p_k(x)] + \sqrt{\frac{\eta_k}{\mu_k}} q_k(\theta)\right)$$
$$= \left(\frac{1}{2} + \frac{1}{2}\sqrt{\eta_k \mu_k}\right)\left(p_k(x) + \sqrt{\frac{\eta_k}{\mu_k}} q_k(\theta)\right).$$

Similarly,

$$K\left(p_k - \sqrt{\frac{\eta_k}{\mu_k}} q_k\right)(x, \theta)$$
$$= \left(\frac{1}{2} - \frac{1}{2}\sqrt{\eta_k \mu_k}\right)\left(p_k(x) - \sqrt{\frac{\eta_k}{\mu_k}} q_k(\theta)\right).$$

Suppose first that $c < \infty$. For $k \geq c$, Lemma A2 shows $E_x q_k(\theta) = 0$ for all $x$. Thus $Kq_k(x,\theta) = \frac{1}{2} q_k(\theta)$. Further

$$\text{span}\bigg\{ p_k(x) \pm \sqrt{\frac{\eta_k}{\mu_k}}\, q_k(\theta)$$
$$0 \leq k < c,\ q_k(\theta) \quad c \leq k < \infty \bigg\}$$
$$= \text{span}\{p_k(x)\ 0 \leq k < c,\ q_k(\theta),\ 0 \leq k < \infty\}$$
$$= L^2(m) + L^2(\pi).$$

It follows that $K$ is diagonalizable with eigenvalues/eigenvectors

$$\frac{1}{2} \pm \frac{1}{2}\sqrt{\mu_k \eta_k}, \quad p_k(x) \pm \sqrt{\frac{\eta_k}{\mu_k}} q_k(\theta) \quad \text{for } 0 \leq k < c,$$
$$\frac{1}{2}, \quad q_k(\theta) \quad \text{for } c \leq k < \infty,$$

and $Kg = 0$ for $g \in (L^2(m) + L^2(\pi))^\perp$.

Suppose next that $c = \infty$; then $K$ is diagonalizable with eigenvalues/eigenfunctions

$$\left(\frac{1}{2} \pm \sqrt{\eta_k \mu_k}\right), \quad p_k(x) \pm \sqrt{\frac{\eta_k}{\mu_k}}\, q_k(\theta), \quad 0 \leq k < \infty.$$

Again, span $\{p_k(x) \pm \sqrt{\frac{\eta_k}{\mu_k}}\, q_k(\theta)\ \ 0 \leq k < c\}$ = span $\{p_k(x), q_k(\theta)\} = L^2(m) + L^2(\pi)$ and $Kg = 0$ for $g \in (L^2(m) + L^2(\pi))^\perp$. This completes the proof of (c). $\square$

## ACKNOWLEDGMENTS

We thank Jinho Baik, Alexi Borodin, Onno Boxma, Vlodic Bryc, Robert Griffiths, Len Gross, Jim Hobert, Susan Holmes, Mourad Ismail, Tom Koornwinder, Angelo Koudou, Christian Krattenthaler, Laurent Miclo, Grigori Olshanski, Dennis Pommeret, Mizan Rahman, Dennis Stanton and Hua Zhou for their enthusiastic help. We particularly thank Gerard Letac for telling us about Lancaster families and for his seminal work on these topics over the years. Finally, we thank two anonymous referees and an editor for their detailed and helpful comments.

Research supported in part by NSF Grants DMS-01-02126 and DMS-06-03886.